\documentclass[10pt]{emulateapj}
\usepackage{apjfonts}
\usepackage{psfig}
\usepackage{epsfig}

\newcommand\puncspace{\ifmmode\,\else{\ifcat.\C{\if.\C\else\if,\C\else\if?\C\else%
\if:\C\else\if;\C\else\if-\C\else\if)\C\else\if/\C\else\if]\C\else\if'\C%
\else\space\fi\fi\fi\fi\fi\fi\fi\fi\fi\fi}%
\else\if\empty\C\else\if\space\C\else\space\fi\fi\fi}\fi}
\newcommand\SP{\let\\=\empty\futurelet\C\puncspace}
\font\smfont=cmr9
\newcommand\Halpha{{H$\alpha$}\SP}                     
\def\Hbeta{{H$\beta$}\SP}                              
\def\Hgamma{{H$\gamma$}\SP}                            
\def\Hdelta{{H$\delta$}\SP}                            
\newcommand\I{\kern.2em{\smfont I}\SP}                 
\newcommand\II{\kern.2em{\smfont II}\SP}               
\newcommand\fnii{\mbox{\rm [N\II]}\SP}                 
\newcommand\foii{\mbox{\rm [O\II]}\SP}                 
\newcommand\HI{\mbox{\rm H\I}\SP}                      
\newcommand\HII{\mbox{\rm H\II}\SP}                    
\newcommand\Ib  {$I$-band\SP}                          
\newcommand\Bb  {$B$-band\SP}                          
\def\BI  {$B-I$\SP}                                    
\newcommand\cf{\textit{cf.} }                          
\newcommand\eg{\textit{e.g.}, }                        
\newcommand\etal{\textit{et al.} }                     
\newcommand\ie{\textit{i.e.} }                         
\newcommand\pone  {Paper I}                            
\newcommand\pthree{Paper III}                          
\newcommand\kms {km~s$^{-1}$\SP}                       
\newcommand\hMpc{h$^{-1}$~Mpc\SP}                      
\newcommand\hkpc{h$^{-1}$~kpc\SP}                      
\newcommand\NGC {NGC\kern.33em}                        
\newcommand\UGC {UGC\kern.33em}                        
\newcommand\AGC {AGC\kern.33em}                        
\newcommand\CGCG{CGCG\kern.33em}                       
\newcommand\IC  {IC\kern.33em}                         
\newcommand\aas {A\&AS}                                
\newcommand\mmras {MmRAS}                              
\newcommand\iaus  {in IAU Symp. }                      

\shorttitle{Properties of Cluster Spirals II.}
\shortauthors{Vogt et al.}

\slugcomment{Accepted by AJ 2004 February 23}

\begin{document} 

\title{M/L, \Halpha Rotation Curves, and \HI Gas Measurements     \break
       for 329 Nearby Cluster and Field Spirals: II. Evidence for Galaxy Infall}

\author{Nicole P. Vogt\altaffilmark{1,2}}
\affil{Department of Astronomy, New Mexico State University, Las Cruces, NM 88003}
\email{nicole@nmsu.edu}
\and
\author{Martha P. Haynes,\altaffilmark{3} Riccardo Giovanelli,\altaffilmark{3}
and Terry Herter}
\affil{Center for Radiophysics and Space Research, Cornell University, 
Ithaca, NY 14853}
\email{haynes, riccardo, and herter@astro.cornell.edu}

\altaffiltext{1}{Formerly at: Institute of Astronomy, University of Cambridge, Cambridge, 
CB3$-$0HA, UK}
\altaffiltext{2}{Formerly at: Center for Radiophysics and Space Research, 
Cornell University, Ithaca, NY 14853}
\altaffiltext{3}{National Astronomy and Ionosphere Center;  NAIC is operated 
by Cornell University under a cooperative agreement with the National Science 
Foundation.}

\begin{abstract}
We have conducted a study of optical and \HI properties of spiral galaxies
(size, luminosity, \Halpha flux distribution, circular velocity, \HI gas mass)
to explore the role of gas stripping as a driver of morphological evolution in
clusters.
We find a strong correlation between the spiral and S0 fractions within
clusters, and the spiral fraction scales tightly with cluster X-ray gas
luminosity.  We explore young star formation and identify spirals that are (1)
{\it asymmetric}, with truncated \Halpha emission and \HI gas reservoirs on
the leading edge of the disk, on a first pass through the dense intracluster
medium in the cores of rich clusters; (2) strongly \HI deficient and {\it
stripped}, with star formation confined to the inner 5 \hkpc and 3 disk scale
lengths; (3) reddened, extremely \HI deficient and {\it quenched}, where star
formation has been halted across the entire disk.  We propose that these
spirals are in successive stages of morphological transformation, between
infalling field spirals and cluster S0s, and that the process which acts to
remove the \HI gas reservoir suppresses new star formation on a similarly fast
timescale.
These data suggest that gas stripping plays a significant role in
morphological transformation and rapid truncation of star formation across the
disk.  \end{abstract}


\renewcommand{\floatpagefraction}{0.70}
\section{Introduction}

Some of the fundamental questions of galaxy formation are the following: Can
we reconcile the observed galaxy populations at high redshifts with those
found today in the local universe?  Are the proposed mechanisms of interaction
between galaxies and other galaxies or clusters sufficient to explain
observations of galaxies within clusters and groups?  Can we similarly explain
the evolutionary history of field galaxies without invoking separate formation
scenarios?

Studies of distant galaxies have established that there is significant
evolution in cluster galaxy populations from redshifts z = 0.5 to the present.
Butcher \& Oemler (1978) first noted the excess of blue galaxies in the cores
of rich clusters at redshifts as low as z $\sim$ 0.2; many studies have since
verified and extended this so--called ``Butcher--Oemler effect''.  Parallel
studies established that the fraction of spiral and S0s varied inversely
within rich clusters, with the S0 fraction dropping by factors of 2-3 from
local levels by redshifts z $\sim$ 0.5 (Couch \etal 1994; Dressler \etal 1997;
van Dokkum \etal 1998; Fasano \etal 2000).  In contrast, the relatively
constant numbers and tightly constrained colors of the cluster elliptical
population indicated that it was very stable, with star formation concentrated
at redshifts z $\ge$ 3 (Ellis \etal 1997, though van Dokkum \etal 1999, 2000
suggest that progenitor bias may elevate the apparent formation redshift).
Finally, optical spectra identified numerous poststarburst galaxies within
clusters (Dressler \& Gunn 1983, Couch \& Sharples 1987), a suppressed star
formation rate for spiral types relative to the field (Balogh \etal 1998), and
found evidence of an infalling spiral population out to redshifts z~$\sim$ 0.4
(Poggianti \etal 1999; Kodama \& Bower 2001).

In summary, these intermediate redshift studies support a well-known class of
scenarios (\eg Spitzer \& Baade 1951; Melnick \& Sargent 1977 for early
discussion) wherein rich clusters are continuously rejuvenated by infalling
field spirals (Balogh, Navarro, \& Morris, 2000); star formation is disrupted
and morphology is slowly transformed, some turning into the S0 population of
today (Jones, Smail, \& Couch 2000; Kodama \& Smail 2001; though see Andreon
1998).

The observational evidence obtained from studying local clusters and groups is
more ambiguous.  Dressler (1980) argued strongly against the formation of S0s
through transformation of spirals, citing their presence in low density, cool
regions, the uniformity of the density-morphology relation in both relaxed and
unrelaxed clusters, and the enhanced luminosity of S0 bulges relative to that
of the complete spiral population.  Solanes \& Salvador-Sol\'e (1992, and
references therein) addressed the first two points by proposing that an
initial correlation between density and bulge mass occurred during protogalaxy
formation, leading to early bulges in pockets of initial overdensities.  This
innate bias could explain the continued morphology-density relation in low
density regions, but would be erased from high density regions during vigorous
stages of cluster formation.  Accreted, gas stripped spirals could gradually
build up the core S0 population afterward, thus restoring the
morphology-density relation.  Pfenniger (1993) addressed the third point with
simulations that suggested that S0 bulges could be supplemented after disk
formation by a burst of star formation, fueled by gas funneled to the nucleus
via a short-lived bar.  This mechanism would also explain the high metallicity
gradients observed in the inner regions of bulges (Fisher, Franx, \&
Illingworth 1996).

The combination of low and intermediate redshift observations indicates that
morphological transformation is significant within clusters, but there is much
to be determined regarding the details of how it occurs.  Three broad types of
physical mechanisms have been proposed to model the transformation of spiral
galaxies in dense environments: galaxy-galaxy interactions, tidal forces, and
gas stripping.  Galaxy-galaxy interactions are most efficient within groups,
where relative velocities are low (Zabludoff \& Mulchaey 1998; Mulchaey \&
Zabludoff 1998; Ghigna \etal 1998), and tidal forces (Toomre \& Toomre 1972;
Moore \etal 1996; Mihos, McGaugh, \& de Blok 1997) are most effective on small
spirals (though see Gnedin 2003a, 2003b).  In spite of these factors, recent
large-scale surveys (Lewis \etal 2002, Nichol \etal 2002) have found clear
evidence of the significance of such moderate physical processes in
suppressing star formation, particularly in galaxies located beyond cluster
cores, though often at a extreme cost in morphological transformation (\eg
harassment conversion into dwarf spheroidals).

Gas stripping mechanisms, in contrast, are extremely efficient at gas removal,
while preserving the large-scale structure of spiral galaxy disks.  They allow
for a gradual fading of stellar populations, as \HII regions diminish in
intensity and surface brightness simultaneously decreases {\it and} becomes
more uniform across the disk.  Stripping can take on several forms, including
absorption of a hot gas outer envelope (Larson, Tinsley, \& Caldwell 1980),
ram pressure sweeping (Gunn \& Gott 1972), evaporation via turbulent mixing
and heat conduction (Cowie \& Songaila 1977), and turbulent viscous stripping
(Nulson 1982).  Hardware limitations (Steinmetz \& M\"{u}ller 1993; Steinmetz
1996) have precluded incorporating the effects into N-body simulations with
gas dynamics at adequate resolution and coverage (time steps and size scales),
though recent models of ram pressure stripping of individual Virgo spirals
have been quite successful (Abadi, Moore \& Bower 1999; Vollmer \etal 1999,
2000, 2001a, 2001b; Vollmer 2003; Vollmer \& Huchtmeier 2003).

Observational studies of the \HI gas dynamics (\cf Giovanelli \& Haynes 1985;
Magri \etal 1988; Solanes \etal 2001) have found a strong correlation between
\HI deficiency and clustercentric radius throughout a range of local clusters,
strongest in early type spirals (Dressler 1986, though note Koopmann \& Kenney
1998), and that while \HI deficiency can extend as far out as 3 \hMpc most gas
stripping occurs well within cluster cores, within galaxies on preferentially
radial orbits (Solanes \etal 2001).  Cayatte \etal (1990, 1994) obtained
aperture synthesis \HI maps of Virgo cluster galaxies and found signatures of
ram pressure and of viscous stripping in separate populations, distinguished
by the ratio of \HI to optical diameters, while Bravo-Alfaro \etal (1997,
2000, 2001) have used a similar technique to identify first pass spirals
falling into and out of the core of Coma and around A262.

An alternate technique is to examine the properties of \Halpha rotation curves
of local cluster spirals.  Early pioneering studies (Rubin, Whitmore, \& Ford
1988; Whitmore, Forbes, \& Rubin 1988; Forbes \& Whitmore 1989) suggested that
velocity profiles declined at large optical radii more in cluster spirals than
in the field, implying gas stripping or diminished halos, but this result has
not been supported by later observations (Distefano \etal 1990; Amram \etal
1993; Sperandio \etal 1995; Vogt 1995).  A recent, large-scale study of 510
rotation curves (Dale \etal 2001) found instead weak trends in the extent and
asymmetry of the \Halpha flux with clustercentric radius, but the strong
selection bias towards late type spirals (Dale \etal 1999) with strong,
extended \Halpha or \fnii 6584\AA\ emission constrains the application of this
result to the general cluster spiral population.

In this paper, we identify and explore a population of infalling spirals in a
large sample of spiral galaxies, 296 selected from 18 nearby clusters and 33
isolated field galaxies observed for comparative purposes.  The current
program integrates both optical and \HI observations and is thus sensitive to
both gas depletion and star formation suppression.  The targeted galaxies are
spread over a wide range of environments, covering three orders of magnitude
in cluster X-ray luminosity and containing galaxies located throughout the
clusters from rich cores out to sparsely populated outer envelopes.  We have
obtained \Halpha rotation curves to trace the stellar disk kinematics within
the potential at high resolution and to explore the strength of current star
formation, \HI line profiles to map the overall distribution and strength of
\HI gas, and \Ib imaging to study the distribution of light in the underlying,
older stellar population.  The sample contains spirals of all types, and is
unbiased by the strength of flux from \HII regions or by \HI gas detection.
This paper is a companion to Vogt, Haynes, Herter \& Giovanelli (2004a,
\pone), which details the observations and reduction of the data set, and to
Vogt, Haynes, Herter \& Giovanelli (2004b, \pthree), which explores changes in
the fundamental parameters (size, luminosity, and mass) and star formation
properties of spiral galaxies as a function of the cluster environment.

\section{Description of Clusters}

\subsection{Characterization}

\begin{table*} [htbp]
  \caption{Cluster Properties}
  \begin{center}
  \begin{tabular} {l c c r r l c l l r@{:}r@{:}l r@{:}r@{:}l r} 
  \tableline
  \tableline
  Cluster   & kT\tablenotemark{a} & log L$_x$\tablenotemark{b}  & $\sigma$\tablenotemark{c} &  V$_{pec}$\tablenotemark{d} & 
  \multicolumn{4}{c}{Cluster Classification\tablenotemark{e}} & \multicolumn{6}{c}{Membership\tablenotemark{f}} & Ref. \\
    & (keV) & (erg s$^{-1}$) & \multicolumn{2}{c}{(km s$^{-1}$)} & 
      J-F   & R              & B-M           & R-S & \multicolumn{6}{c}{(E:S0:Sp)} \\
  \multicolumn{1}{c}{(1)}  & \multicolumn{1}{c}{(2)}  & \multicolumn{1}{c}{(3)}  & \multicolumn{1}{c}{(4)}  & \multicolumn{1}{c}{(5)}  & 
  \multicolumn{1}{c}{(6)}  & \multicolumn{1}{c}{(7)}  & \multicolumn{1}{c}{(8)}  & \multicolumn{1}{c}{(9)}  & \multicolumn{3}{c}{(10)} & 
  \multicolumn{3}{c}{(11)} & \multicolumn{1}{c}{(12)} \\
  \tableline 
  A1656     &  8.3  &  45.10  &  997 &     170 & evolved nXD & 2 & II       &  B     &  28&50&22 &  35&47&18 & 1,4,13  \\
  A426      &  6.2  &  45.36  & 1307 &    -364 & evolved XD  & 2 & II$-$III &  L     &  30&30&41 &  48&45&07 & 1,5,14  \\
  A2199     &  4.7  &  44.81  &  823 &    -235 & int. XD     & 2 & I        & cD$_p$ &  23&31&46 &  35&41&24 & 1,6,13  \\
  A2147     &  4.4  &  44.58  &  821 &     303 & early XD    & 1 & III      &  F     &  11&24&66 &  27&31&42 & 1,6,15  \\
  A2063     &  4.1  &  44.47  &  626 &     680 & int. XD     & 1 & II$-$III & cD$_p$ &  13&41&46 &  38&13&49 & 1,6,16  \\
  A2151     &  3.8  &  44.00  &  705 &     312 & early XD    & 2 & III      &  F     &  13&19&68 &  14&35&51 & 1,7,13  \\
  A1367     &  3.7  &  44.23  &  802 &      43 & early nXD   & 2 & II$-$III &  F     &  11&26&63 &  17&40&43 & 1,6,13  \\
  A2634     &  3.4  &  44.12  &  661 &     -82 & early nXD   & 1 & I$-$II   & cD     &  17&42&41 &  17&47&36 & 1,8,16  \\
  A539      &  3.0  &  43.80  &  701 &    -277 & nXD         & 1 & III      &  F     &  07&42&51 &  19&53&28 & 1,9,13  \\
  A262      &  2.4  &  43.93  &  575 &     -32 & early XD    & 0 & III      &  I     &  09&32&59 &  17&36&47 & 1,10,14 \\
  A400      &  2.1  &  43.82  &  621 &    -250 & int. XD     & 1 & II$-$III &  I     &  09&34&58 &  15&56&29 & 1,6,13  \\
  A2152     &  2.1  &  43.49  &  715 & \nodata & nXD         & 1 & III      &  I     &  12&28&60 &  38&29&33 & 2,6,15  \\
  A2666     & (1.7) & (42.00) &  476 &   -156  & \nodata     & 0 & I        & cD$_p$ &  15&32&52 &  20&37&43 & 2,11,14 \\
  A2197     &  1.6  &  43.08  &  550 &    -282 & nXD         & 1 & II       &  L     &  15&03&82 &  19&36&45 & 2,6,13  \\
  N507      & (1.6) & \nodata &  444 &     242 & nXD         & 0 & III      &  F     &  11&30&59 &    &  &   & 2,10    \\
  A779      &  1.5  &  42.95  &  503 &    -100 & \nodata     & 0 & II       & cD$_p$ &  03&10&86 &  04&12&85 & 2,6     \\
  A2162     & (0.9) &  42.95  &  323 & \nodata & nXD         & 0 & II$-$III &  I     &  11&05&84 &  11&06&83 & 2,6     \\
  Cancer    & (0.9) & $<42.30$\hspace{0.1in} & 317 & 250 & nXD & 0 & III    &  I     &  18&12&70 &  11&18&71 & 3,12    \\
  \tableline 
  \multicolumn{16}{l}{\hspace{0.05truein} $^a$X-ray temperatures from Wu, Fang, \& Xu (1998), and $^b$bolometric luminosity;  parenthesized } \\
  \multicolumn{16}{l}{values derived from velocity dispersion.} \\
  \multicolumn{16}{l}{\hspace{0.05truein} $^b$Central velocity dispersion} \\
  \multicolumn{16}{l}{\hspace{0.05truein} $^c$Peculiar velocity, taken from Giovanelli {\textit{et al.}} (1997); derived in consistent fashion for } \\
  \multicolumn{16}{l}{additional clusters.} \\
  \multicolumn{16}{l}{\hspace{0.05truein} $^d$Rankings under Jones-Forman, Abell richness, Bautz-Morgan, and Rood-Sastry classification } \\
  \multicolumn{16}{l}{systems.} \\
  \multicolumn{16}{l}{\hspace{0.05truein} $^e$Morphological ratio for members out to 2 $h^{-1}$ Mpc, and from the literature (drawn from inner } \\
  \multicolumn{16}{l}{1.5 $h^{-1}$ Mpc).} \\
  \multicolumn{16}{l}{X-ray data taken from (1) David \etal (1993), (2) Abramopoulos \& Ku (1983, (3) Giovanelli \& } \\
  \multicolumn{16}{l}{Haynes (1985).  Velocity dispersions taken from (4) Kent \& Gunn (1982), (5) Kent \& Sargent } \\
  \multicolumn{16}{l}{(1983), (6) Zabludoff \etal 1993, (7) Bird, Dickey, \& Salpeter (1993), (8) Scodeggio \etal (1994), } \\
  \multicolumn{16}{l}{(9) Ostriker \etal (1988), (10) Sakai, Giovanelli, \& Wegner (1994), (11) Struble \& Ftaclas (1994), } \\
  \multicolumn{16}{l}{(12) Bothun \etal (1983).  Morphological ratios taken from (13) Oemler (1974), (14) Melnick \& } \\
  \multicolumn{16}{l}{Sargent (1977), (15) Tarenghi \etal (1980), (16) Dressler (1980b).} \\
  \end{tabular}
  \end{center}
  \label{tab:cl_prop1}
\end{table*}

As discussed in \pone, we have selected a set of local clusters and groups
which span a wide range of environments, parameterized by X-ray luminosity and
velocity dispersion, richness, substructure, and spiral fraction.
Table~\ref{tab:cl_prop1} lists the complete set, hereafter referred to as the
cluster sample, ordered by X-ray temperature, and compares them under several
forms of cluster classification.  We include two estimates of the
morphological fraction, one taken from the literature, and a second determined
from all galaxies with measured positions and redshifts within 2 \hMpc of the
cluster center. The latter measure incorporates all cataloged galaxies,
referred to as the ``parent sample'', encompassing the same volume and to the
same depth, as our spiral galaxy study. Galaxies with measured redshifts
within 6$^{\circ}$\ (5 to 10 \hMpc) of each cluster center were selected; the
assembled galaxies range from 10 to 22 in \Bb magnitudes, peaking at 15.5
magnitudes.  As discussed below, a subsample comprised of $\sim$ 300 spiral
galaxies inclined more than 30$^{\circ}$\ from face-on orientation was then
drawn from the parent sample, to serve as targets for a detailed dynamical
study of the process of spiral infall.

The initial cluster galaxy sample was drawn from the 1994 version of the
private database of R.G. and M.P.H. known as the \AGC, including objects
contained within the UGC and CGCG galaxies, objects included in the cluster
sample of Dressler (1980) and objects identified by eye examination of the
POSS prints.

We supplemented these data with those for additional galaxies archived within
the NASA/IPAC Extragalactic Database\footnote{The NASA/IPAC Extragalactic
Database is operated by the Jet Propulsion Laboratory, California Institute of
Technology, under contract with the National Aeronautics and Space
Administration} (NED), and the results of numerous local redshift surveys.
The combined number of galaxies with measured redshifts within $4\sigma$ and 2
\hMpc for each cluster ranges from 35 for the poorest cluster A2162 to 583 for
the rich A1656, with a median value of 121.  This parent sample starts to
become significantly incomplete at \Bb magnitudes ranging from $-$18.5 to
$-$19.5 over the redshift range of the sample ($z$ = 0.016 to 0.037).
However, data from the well-studied cluster A1656 (Coma) extend more than a
full magnitude deeper.  The sample is comprised of clusters with well-studied
dynamics, and thus while it is not strictly complete in magnitude, size, or
volume, it is assumed to be fairly complete to roughly two magnitudes below
$L^*$ for cluster members within 3 \hMpc of the cluster centers.

\subsection{Distribution of Parent Sample of Galaxies}

\begin{figure*} [htbp]
  \begin{center}\epsfig{file=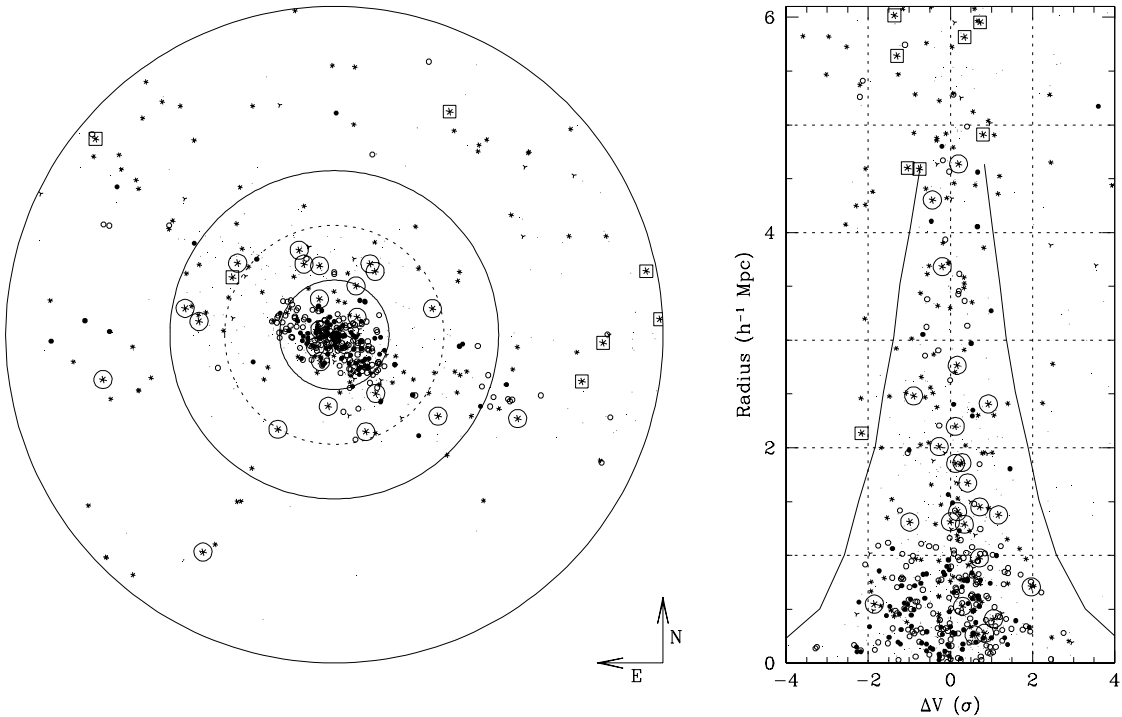,width=6.0truein}\end{center}
  \caption[A1656 Cluster Environs]  
{Distribution of galaxies within 6 \hMpc and 4$\sigma$ of hot cluster A1656
(Coma).  Galaxies are drawn as ellipticals (filled circles), S0s (open
circles), spirals (asterisks), irregulars (skeletal triangles) and untyped
(small dots).  Those within our dynamical sample are surrounded by circles
(cluster members), squares (galaxies associated with the cluster), or diamonds
(foreground and background galaxies, none within A1656).  The rectangular plot
shows the radial velocity distribution, in units of the cluster velocity dispersion; the
membership caustic is taken from Kent \& Gunn (1982).  The round plot shows
the spatial distribution of galaxies around the cluster, with circles drawn at
1, 2, 3, and 6 \hMpc.  It extends to show ten galaxies at radii well beyond 3
\hMpc; we have sampled the extended supercluster structure more here than
around the other clusters.  Though beyond the cluster proper, these galaxies
fall at the same redshift and are part of a larger unbound structure.}
  \label{fig:mship1}
\end{figure*}

\begin{figure*} [htbp]
  \begin{center}\epsfig{file=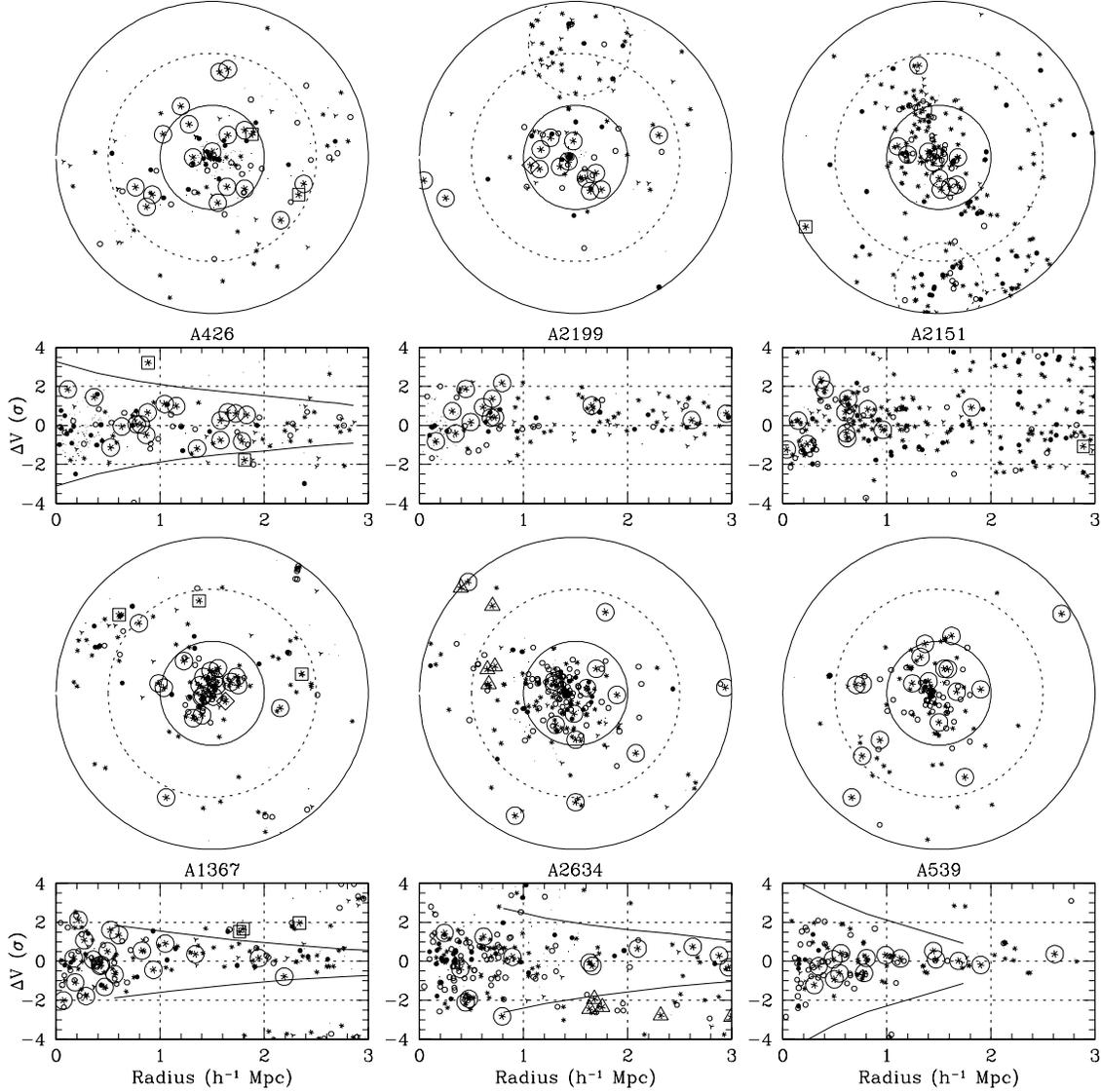,width=6.0truein}\end{center}
  \caption[A426 through A539 Cluster Environs]
{Distribution of galaxies within 3 \hMpc and 4$\sigma$ of hot clusters, with
clusters ordered in decreasing X-ray temperature; symbols are as in
Figure~\ref{fig:mship1}, and members of the infalling foreground group A2634-F
around A2634 are surrounded with triangles.  The round plots show the spatial
distribution of galaxies, with circles drawn at 1, 2, and 3 \hMpc and
neighboring clusters marked with dashed circles at 1 \hMpc.  Membership
caustics are shown for A426 (Kent \& Sargent 1983), A1367 (Giovanelli \etal
1997a), A2634 (Scodeggio \etal 1995), and A539 (Ostriker \etal 1988).
Galaxies were partitioned between A2197 and A2199 along a dividing line at
40$^{\circ}$ 30\arcmin.  Hercules supercluster membership was originally
established by breaking the sample into galaxies north of 17$^{\circ}$\
(A2151), and then galaxies east (A2152) and west (A2147) of 16$^{\circ}$
03\arcmin, and then updated in accordance with the extensive kinematic study
of Barmby \& Huchra (1998).
}
  \label{fig:mship2}
\end{figure*}

\begin{figure*} [htbp]
  \begin{center}\epsfig{file=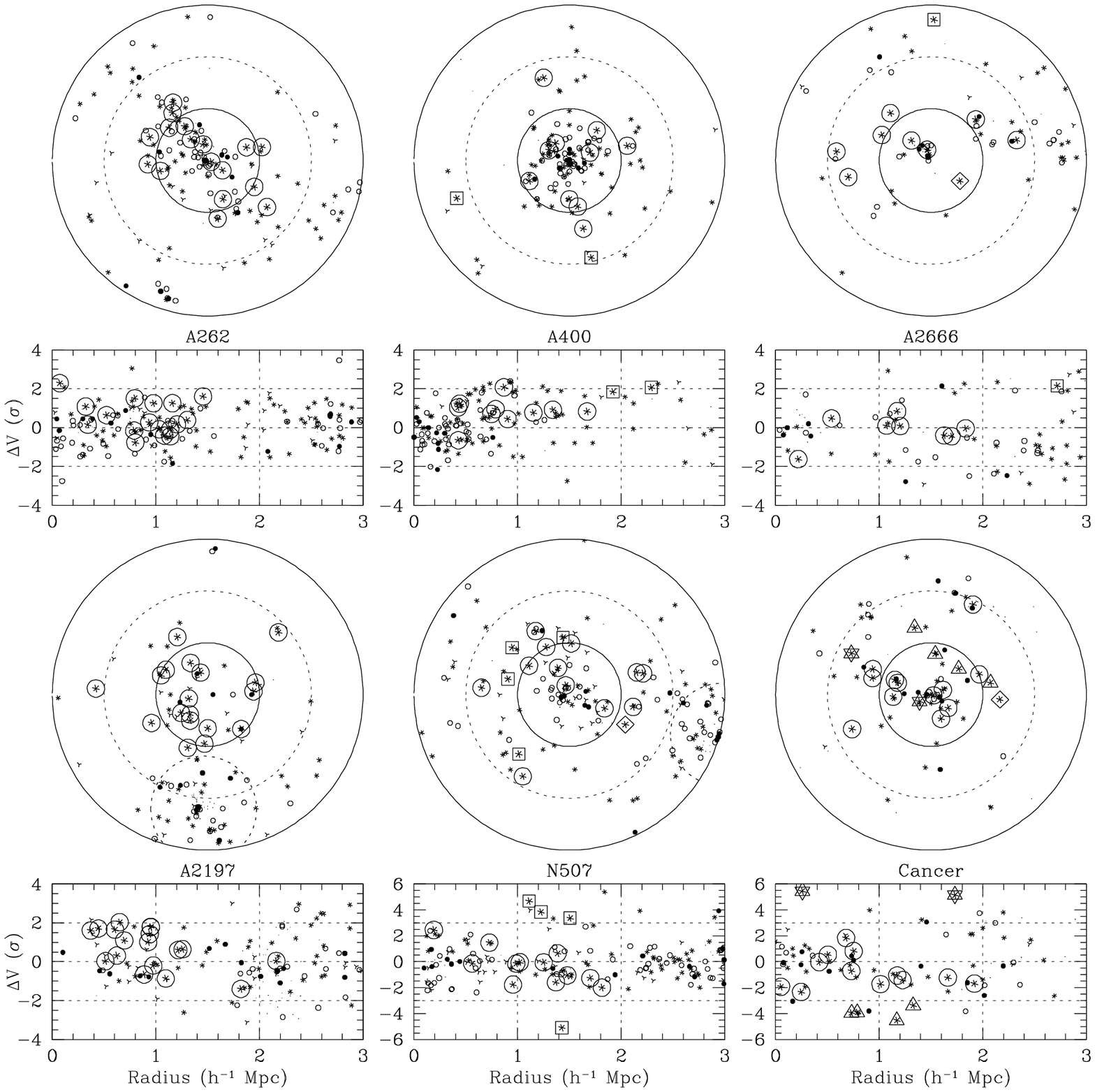,width=6.0truein}\end{center}
  \caption[A262 through Cancer Cluster Environs]
{Distribution of galaxies within 3 \hMpc of cold clusters, with clusters
ordered in decreasing X-ray temperature; symbols as in Figure~\ref{fig:mship1}.
The round plots show the spatial distribution of galaxies, with circles drawn
at 1, 2, and 3 \hMpc and neighboring clusters marked with dashed circles at 1
\hMpc.  The Cancer and \NGC507 cluster radial velocity diagrams are extended to 
6$\sigma$ to show infall and the main subclumps within Cancer (as defined in
Bothum \etal 1983); galaxies within our sample are surrounded by circles
(clump A members), stars (clump B or C members), triangles (clump D members),
or diamonds (background galaxy).}
  \label{fig:mship3}
\end{figure*}

Figures~\ref{fig:mship1} through \ref{fig:mship3} show the distribution of
galaxies within the parent sample around each cluster, on the sky and in
radial velocity space.  The clusters have been sorted in order of decreasing
X-ray temperature; for the purposes of discussion, we divide the sample
between {\it hot} and {\it cold} clusters, at kT$_{gas} = 3$ eV.  Figure 2
shows the distributions for hot clusters, while Figure 3 shows similar
displays for the cold ones; the Coma cluster is illustrated individually in
Figure 1.  Galaxies are restricted to the range 4$\sigma$ (velocity
dispersion) about each cluster.  This criterion is relaxed to 6$\sigma$ for
\NGC507 and Cancer to show the multiple subclumps within each group, as we
have chosen to use the velocity dispersion of the dominant subgroup (listed in
Table~\ref{tab:cl_prop1}) rather than of the entire population.  We selected
spirals within 2 \hMpc at the highest priority for our dynamical study,
starting at the cluster centers and spiraling outwards.  Despite this
weighting scheme, we have observed no spirals within 200 \hkpc and 1$\sigma$
of any of the cluster cores.  This is not surprising, particularly for the hot
clusters where many of the cluster centers are defined as the position of a cD
galaxy.  The cD halo alone could extend this far, and any spiral drawn this
near would be subsumed by tidal forces.

Galaxies have been divided into ellipticals, S0s, and spirals to illuminate
morphological segregation, as well as the enhanced overall density of galaxies
in the cores.  Targets of our dynamical study are further identified by larger
concentric symbols according to their relationship to the cluster: true
cluster members (circles), galaxies associated with the cluster potential and
thus infalling (squares) and foreground and background galaxies (diamonds).
The designation of true cluster member has been reserved for galaxies within
the main envelope of the cluster, while associated galaxies (either at rest
with respect to the cluster at large radii, or, less commonly, in the inner
few Mpc with a substantial velocity offset).  The most significant
complication in applying membership criteria lies at radii beyond 1 \hMpc, in
the removal of contamination from neighboring clusters, as discussed in
Figures~\ref{fig:mship2} and \ref{fig:mship3}.  The associated galaxies
identified to be infalling into the cluster potential may also include true
members on extreme radial orbits which have already passed through the center
of the cluster.  True cluster members are assumed to be at rest with respect
to the cluster, while the associated members are taken to be at distances
corresponding to their individual redshifts.  The distinction is most relevant
for galaxies offset in velocity from the cluster center.

\begin{figure} [htbp]
  \begin{center}\epsfig{file=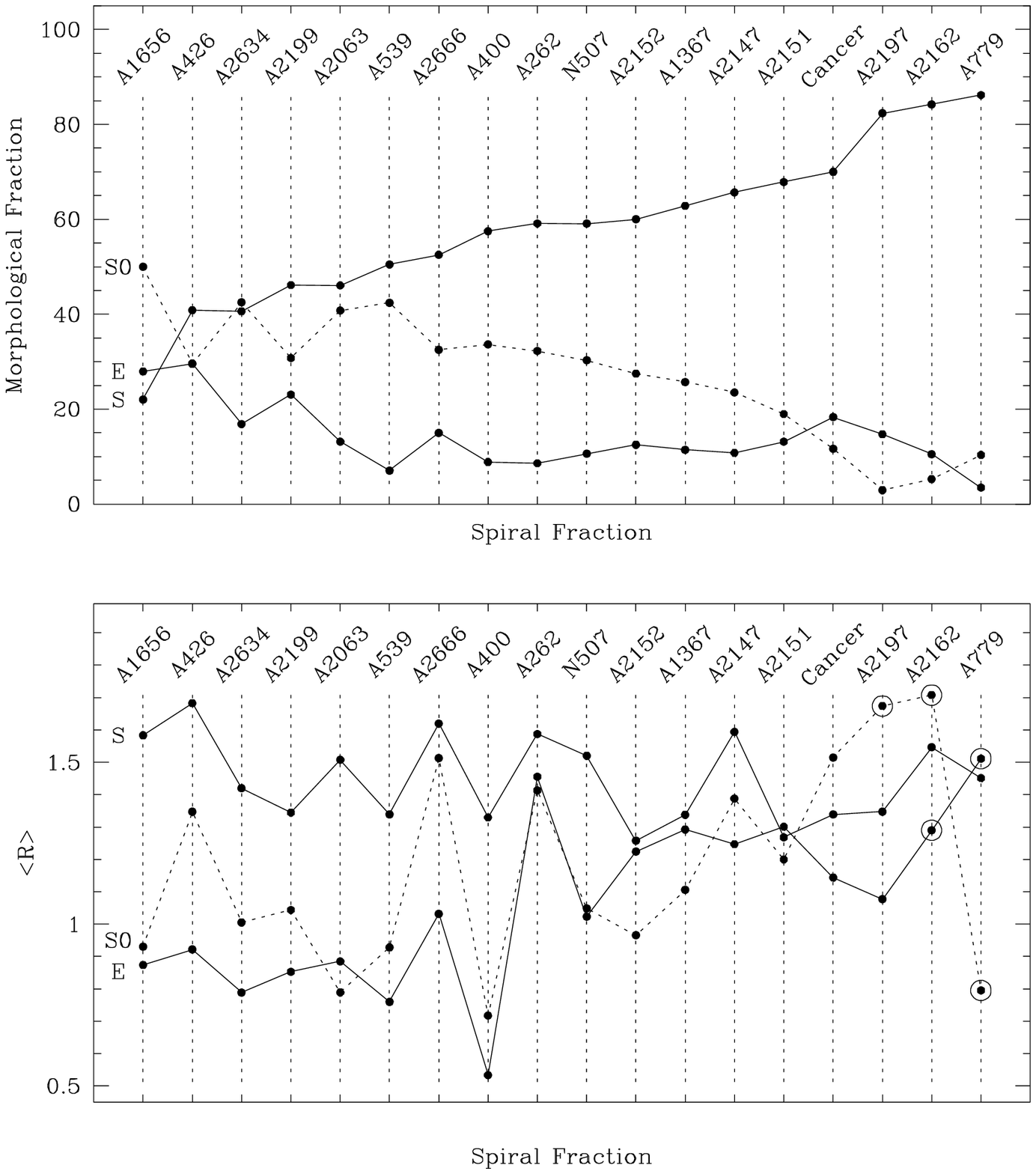,width=3.5truein}\end{center}
  \caption[Morphological fraction and R mean values]
{Distribution of relative morphological fractions {\bf (top)} and mean values
of the offset, R, {\bf (bottom)} for ellipticals, S0s, and
spirals within each cluster.  Clusters are ordered by spiral fraction.  Points
are circled when the mean of the offset R was determined from fewer
than five points.  The fraction of ellipticals stays around 15\%, while there
is a clear inverse correlation ($r = -0.92$) between the fraction of spirals
and S0s.  The mean R value for spirals is relatively constant across
the sample.  Ellipticals are distributed similarly to spirals in the poor
clusters, while those in the rich, hot clusters are far more centrally
concentrated.}
  \label{fig:mrp11} 
\end{figure} 

Figure~\ref{fig:mrp11} (top) summarizes the morphological type distribution
across the parent sample, where the clusters have been ordered by increasing
spiral fraction, calculated from all cataloged galaxies with 2 \hMpc.  It
should be noted that the spiral fractions derived from the literature
typically refer to a smaller volume ($<$ 1.5\hMpc), whereas those derived from
our parent galaxy incorporation include a greater contribution from the outer,
typically spiral-richer regions. The general trend is that the spiral fraction
varies inversely with the X-ray temperature of the cluster.  It includes rich
clusters A1367 and A2151 with high spiral fractions, while poor clusters A2666
and \NGC507 have few members but contain a high fraction of S0s.  The high
spiral fraction within A2147, in contrast, is expected, given that 90\% of the
X-ray luminosity is contributed by an AGN rather than by diffuse gas (Ebeling
\etal 1990).  The fraction of ellipticals within the clusters is relatively
constant, hovering around 15\% regardless of X-ray temperature or spiral
fraction, and below 25\% for all but two clusters, and there is a high
correlation ($r = 0.94$) between counts of elliptical and S0 members; note
that this has been observed out to redshifts beyond $z \sim 0.5$ (\cf Dressler
\etal 1997).  We find an equally strong inverse correlation ($r = -0.92$)
between the {\it fraction} of spirals and of S0s which make up the remaining
85\%.  This trend holds across a range of cluster membership algorithms, and
is evident in samples extending out to between 1 and 3 \hMpc, beyond the
virial radius of even the richest clusters where the density drops to $\le 3$
(h$^{-1}$ Mpc)$^2$ galaxies, and in the morphological ratios assembled from
heterogeneous cluster surveys in the literature.

We find a moderately strong correlation ($r = -0.70$) between spiral fraction
and X-ray gas luminosity or temperature (dropping to $r \sim 0.5$ for S0s or
ellipticals), though no significant difference that might allow us to explore
the relationship between intracluster gas density and the cluster potential
(\ie mass) within the clusters.  Edge \& Stewart (1991) have reported a much
stronger correlation ($r = -0.96$ for luminosity, $-0.85$ for kT$_{gas}$) for
a similarly sized sample, restricted to clusters with solid EXOSAT detections
and kT$_{gas} > 2$ eV, but we do not reproduce it within the hotter portion of
our sample.  The stronger correlation, across the entire sample, is in fact
between the X-ray gas luminosity and the temperature ($r = 0.92$).

We define an overall measure, R, of the distance of each galaxy from its host
cluster center by normalizing and combining the radial and velocity offsets.
We equate a velocity offset of 3$\sigma$ with a distance offset of 2 \hMpc,
such that, for radius $r$ in units of \hMpc and velocity $v$ in units of
$\sigma$,
\begin{equation}
  R = \sqrt { {r}^2 + { (\frac{2}{3} v) }^2 } \;\; \mbox{h}^{-1} \mbox{Mpc.}
  \label{eq:r_phase}
\end{equation}
Figure~\ref{fig:mrp11} (bottom) shows the distribution of mean values of R for
each morphological type within the clusters, where we have relaxed our
membership criteria to include galaxies out to 3 \hMpc within 3$\sigma$.  The
spiral population (both early and late types) has a relatively constant mean
of 1.5 \hMpc across the sample, representative of the spiral envelope which
surrounds all of the cluster cores.  Ellipticals match the spiral distribution
in poor clusters, but in the rich clusters the mean falls to a more centrally
concentrated 0.85 \hMpc.  This trend may be paralleled in the S0s, though the
scatter is considerable.  The velocity offset differences among morphological
populations are negligible; this effect is driven by the radial distribution
and the high concentration of ellipticals found in the cores of the rich
clusters.

\begin{figure*} [htbp]
  \begin{center}\epsfig{file=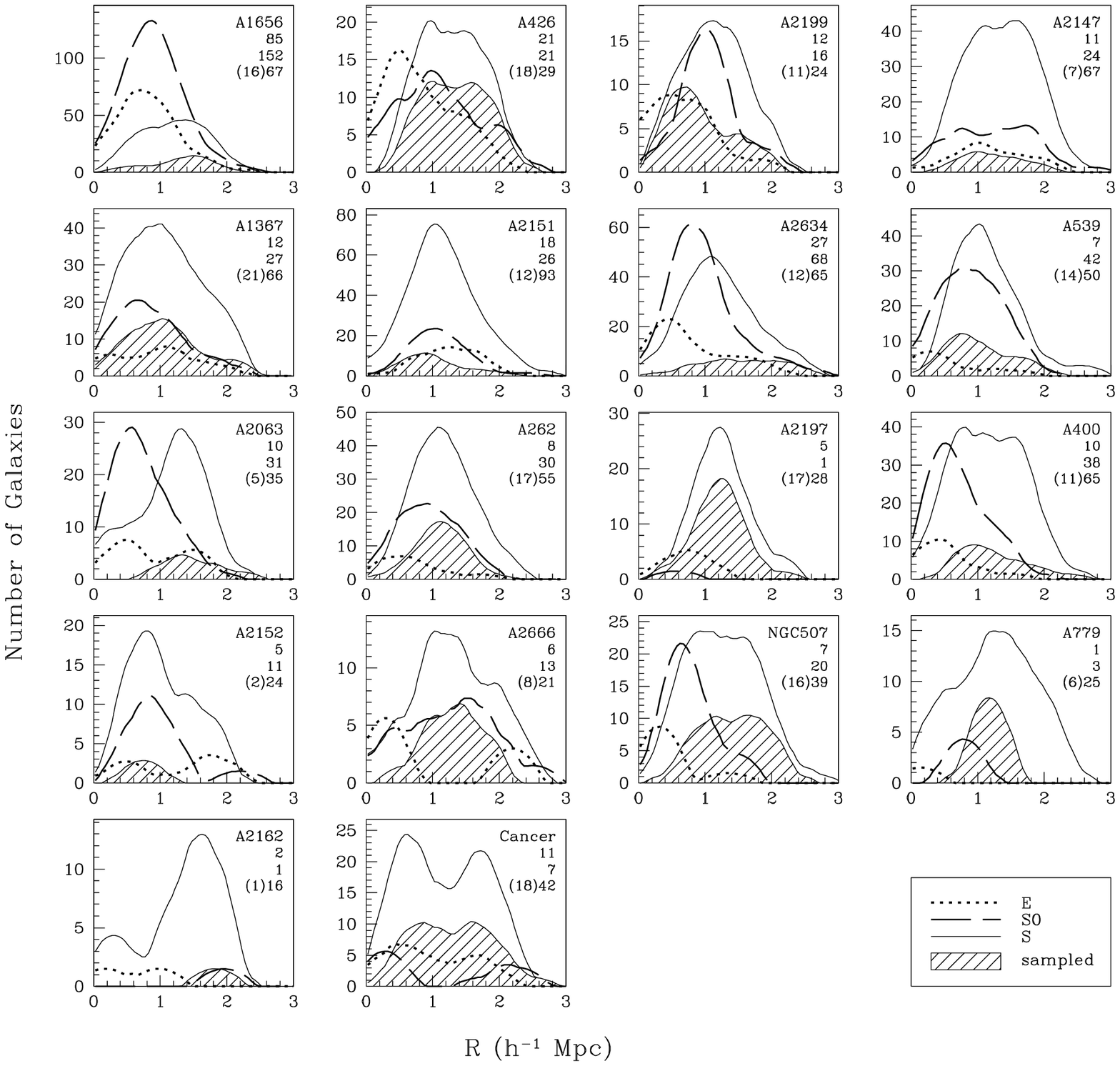,width=6.0truein}\end{center}
  \caption[Cluster R Distributions]
{Distribution of the cluster offset, R, for galaxies of different morphological
types within each cluster, ordered by X-ray temperature.  Galaxies must fall
within the inner $\frac{1}{2}$ \hMpc and $5\sigma$, or 2 \hMpc and 3$\sigma$
(relaxed to 6$\sigma$ for Cancer and \NGC507), of the cluster to be included,
and neighboring cluster contamination has been removed.  The dotted line shows
the distribution of ellipticals, the dashed line the S0s, and the
solid line the spirals.  The hatched areas show the subset of the parent spiral
distribution targeted within our dynamical sample.  The number of ellipticals,
S0s, and spirals is indicated below the cluster name, with the number
of spirals within our dynamical sample in parenthesis.}
  \label{fig:mrp10}
\end{figure*}

Figure~\ref{fig:mrp10} shows the distribution of mean R values for
ellipticals, S0s, and spirals within each cluster, for members within
$\frac{1}{2}$ \hMpc and 5$\sigma$ or within 2 \hMpc and 3$\sigma$, with
neighboring clusters removed.  The plots extend to 3 \hMpc in R, and the bulk
of the cluster members fall within 2 \hMpc.  Galaxies within the sample which
fall beyond these limits (\eg spirals in the outer regions of the cluster
environs, background and foreground galaxies) are thus not included on these
plots, nor in the counts of galaxies of each type listed below the name of
each cluster.  The distribution was determined by applying a fixed-width
kernel density estimate (Silverman 1986) to the raw R values, to achieve
smoother representation than a standard histogram.  An Epanechnikov kernel
(inverted parabola) with a width of 0.5 R was used for the entire data set;
selected by applying least-squares cross-validation to the distribution of
each morphological type within each cluster.

The effects of morphological segregation can be clearly seen in the relative
proportions of early to late type galaxies, ranging from the S0 dominated and
elliptical rich A1656 down to poor clusters such as A779, composed almost
entirely of spirals.  As in Figure~\ref{fig:mrp11}, we observe both rich
clusters with high spiral fractions well within the potential and poor
clusters with a significant number of S0s.  These data will also be used to
evaluate the validity of the subsampling of the parent spiral galaxy
population in the selection of the spirals that comprise our dynamical sample.

\subsection{Distribution of Dynamical Subsample of Galaxies}
\label{subsec:ExamSmp}

For a sample of 329 spiral galaxies, 296 in the vicinity of the clusters
listed in Table~\ref{tab:cl_prop1}, we have obtained \Halpha rotation curves
to trace the stellar disk kinematics and the extent of young star formation,
\HI line profiles to map the overall distribution and strength of \HI gas, and
\Ib imaging to study the distribution of light in the underlying, older
stellar populations.  We have one or more optical spectra for every targeted
galaxy.  However, it was not possible to survey all of the clusters within the
sample to completion because of various observational constraints (\eg adverse
weather conditions, successful acquisition of a photometric optical image and
\HI line profile for each new galaxy with an optical spectrum). We must thus
examine each cluster sample individually to determine whether the observed
galaxies comprise a valid representation of the parent cluster spiral
population.

In the evaluation of sample validity, we examine the final observational
sample according to five criteria. Three refer to the size and extent of the
targets: (1) We must have obtained 15 or more optical spectra in the region of
each cluster, and this sample must not be biased against galaxies undetected
in \HI.  This number includes true cluster members, galaxies associated with
the cluster (\eg infalling spirals on the outskirts), and foreground and
background galaxies; together these are designated as the {\it observed
sample}.  (2) We must have obtained 10 or more optical spectra of true cluster
members within the observed sample.  The distinction between true members,
associated galaxies, and foreground and background galaxies has been made for
each observed galaxy on a case by case basis by examining the parent
distribution of galaxies on the sky and in radial velocity space (see Figures
\ref{fig:mship1} through \ref{fig:mship3}).  (3) If \HI line profiles and
optical imaging have not been obtained for the complete observed sample, the
galaxies for which such data is missing must form an unbiased subset.

The final two criteria contrast the observed subsample for each cluster with
the parent sample of spirals in the region.  (4) We must have observed more
than 10\% of the parent cluster spiral population. (The parent sample is $<$
0.5 magnitudes deeper than the observed sample in completeness; see \pthree\
for an analysis of completeness in the observed dynamical subsample.)  A case
by case study of membership such as that done for the observed sample would
have been very time consuming as there are more than 100 spirals in some of
the clusters, so we have used the same constraints as in
Figure~\ref{fig:mrp10} to define cluster membership for all parent sample
galaxies in the vicinity of each cluster.  The average fraction observed for
the well sampled clusters is 29\%, in line with our restriction to inclination
angles greater than 30$^{\circ}$.  Note, however, the element of
self-fulfilling prophecy, as the observed sample is always completely
represented in the parent population, which in turn is sensitive to the depth
to which the cluster region has been explored in redshift surveys.  (5) The
correlation between R distributions between the observed sample and the parent
cluster spiral population must not be strongly biased, as compared to Monte
Carlo simulations of randomly chosen subsamples of the same size.

The hatched region on each cluster plot in Figure~\ref{fig:mrp10} represents
the spirals within our subsample; we can compare their distribution with that
of the spiral fraction of the parent sample.  We assume that the combined \AGC
and NED data provide a good representation of the complete population of
spirals within the cluster, to within our magnitude limits, and compare the
distribution of R within the dynamical subsample to it.  The shape of the
parent R distribution is frequently mirrored in the subsample, as in the case
of A1656 and A539.

We have computed a correlation function between the R distribution of the
parent and subsampled spirals within each cluster, as a function of R.  The
validity of the measured correlation is highly dependent upon the number of
galaxies within each sample, so it is not enough to measure the correlation
alone.  We have run Monte Carlo trials on each cluster, sampling without
replacement the parent set of spirals to form 100 randomly selected subsamples
the same size as each observed subsample.  We then compare the distribution of
the correlation coefficients between the parent sample and each of the
simulated subsamples, and the correlation for the observed subsample.
Undersampled clusters were identified by a wide variation in the correlation
of the simulated subsamples, and well sampled clusters were examined for signs
of selection bias.

Clusters such as A426 and A262 were not sampled beyond a radius of 2 \hMpc in
the dynamics program, and thus the extreme outer envelope of parent spirals is
undersampled at R $\geq$ 2.5.  This is acceptable for our purposes, given our
focus on the inner 2 \hMpc.  Three other clusters, however, show significant
differences between parent and subsampled R distributions.  The subsample for
A2063 has a markedly different shape from that of the parent spiral
population, particularly in the inner region.  This is because the subsampling
is very incomplete, and highly biased towards galaxies with strong 21~cm line
profiles, in contrast to the other clusters.  The clusters A779, A2147, A2152
and A2162 have not been sampled deeply enough and lack sufficient numbers in
the observed sample.  We thus discard all five from the dynamical program;
they are not shown in Figures~\ref{fig:mship1} through~\ref{fig:mship3}.

The remaining 13 clusters appear to have been well sampled, according to the
above limits.  The first two criteria have been relaxed slightly for A2666.
It lies 4 \hMpc from A2634 in a well sampled region, with many redshifts in
the literature, so we are confident that the parent spiral population is a
good representation of the actual population in the region, down to our
magnitude limits.  Though we have observed only eight true cluster members
they make up 38\% of the spirals in the parent population of A2666, the
expected fraction given our restriction in inclination angle.  We thus assume
that the small size of the observed sample reflects the limitation imposed by
the small actual number of spirals within the cluster.

We lack \HI line profiles for a significant number of galaxies within clusters
A2151, A2197, and A2199, unlike the other well sampled clusters.  Many of the
galaxies in A2197 and A2199 were unreachable from Arecibo, and we did not have
sufficient observing time to survey A2151; the galaxies without \HI line
profiles are not biased significantly (\ie due only to the declination limit
of the Arecibo dish for A2197 and A2199) relative to the observed sample.  We
have taken the precaution of conducting our analysis by including and then
discarding the galaxies within these three clusters, and find no significant
difference in the results.

\section{Direct Evidence for Infall}
\label{subsec:EviTrans} 

The goal of our program is to explore the effects of the cluster environment
on spiral galaxies.  We have focused our efforts on two facets: (1) direct
effects of infall on field spirals on a first pass towards the cluster core;
(2) fundamental differences in the structure of cluster spirals relative to
the field population, which could be caused either by perturbations from
recent (or long-distant) infall, or by initial disk formation in a
circumscribed, over-dense environment (\eg halo truncation).  The combination
of multiwavelenth observations spread across many clusters offers a
complementary approach to higher resolution studies focussed on single
clusters.  We have obtained single-dish \HI gas line profiles, moderate
resolution major axis \Halpha and \fnii optical spectra, and photometric \Ib
images for our sample, augmented with \Bb total magnitudes extracted as
available from a variety of literature sources (primarily the RC3). We begin
by identifying key observables within our data set which suggest a current or
recent disturbance due to the cluster environment.

Albeit less individually illustrative, single--dish \HI line profiles are
observationally cheaper than two-dimensional \HI maps and thus can be sampled
in a wide range of environments as we have done here. The shapes of
single-dish \HI line profiles are sensitive to a number of factors unrelated
to tidal interactions or gas stripping, including the inclusion of small,
gas-rich companions within the telescope beam, high velocity clouds, and warps
in the \HI disk structure. However, the total \HI gas mass is a key tracer of
gas stripping, and, though crude, \HI line profile shapes are still a powerful
secondary indicator when used in conjunction with spatially resolved velocity
profiles (see below).

The \HI deficiency, as discussed in \pone, is a measure of the difference
between the measured \HI mass and that expected for a galaxy of similar
morphology and size. We divide our dynamical sample into an \HI normal
(gas-rich) population and an \HI deficient (gas-poor) population, based on
galaxy type and blue radius R$_b$ correlations determined from a large body of
field spirals (Solanes \etal 1996, 2001), where gas-poor spirals are deficient
in \HI by a factor of 2.5 or more (log \HI$_{def} \ge$ 0.40).  Note that many
cluster spirals as so deficient in \HI gas that any remnant cannot be easily
detected, and the quoted \HI deficiency is an upper limit on the remaining gas
mass based on instrumental sensitivity.  In these cases, we have applied
survival analysis to the observed upper limits (see \pone).

Likewise, maps of the \Halpha as obtained from Fabry--Perot or fiber bundle
techniques provide more detailed pictures of asymmetries but are relatively
expensive to obtain. With their recognized limitations, we use spatially
resolved \Halpha and \fnii optical spectra as primary indicators for
infall-induced distortion.  By evaluating the \Halpha flux distribution along
the major axis rotation curves, we obtain a measure of the young star
formation, and thus molecular gas and \HII region strength across the entire
disk.  Our exposures are deep enough and the slit wide enough (20 - 40
{h$^{-1}$ pc) that the observed signal is spatially continuous rather than a
series of isolated delta functions caused by individual \HII regions entering
the slit.  It serves well as an estimate of the radial flux profile
characteristic of the disk from one side to the other for such inclined
galaxies.

\begin{figure*} [htbp]
  \begin{center}\epsfig{file=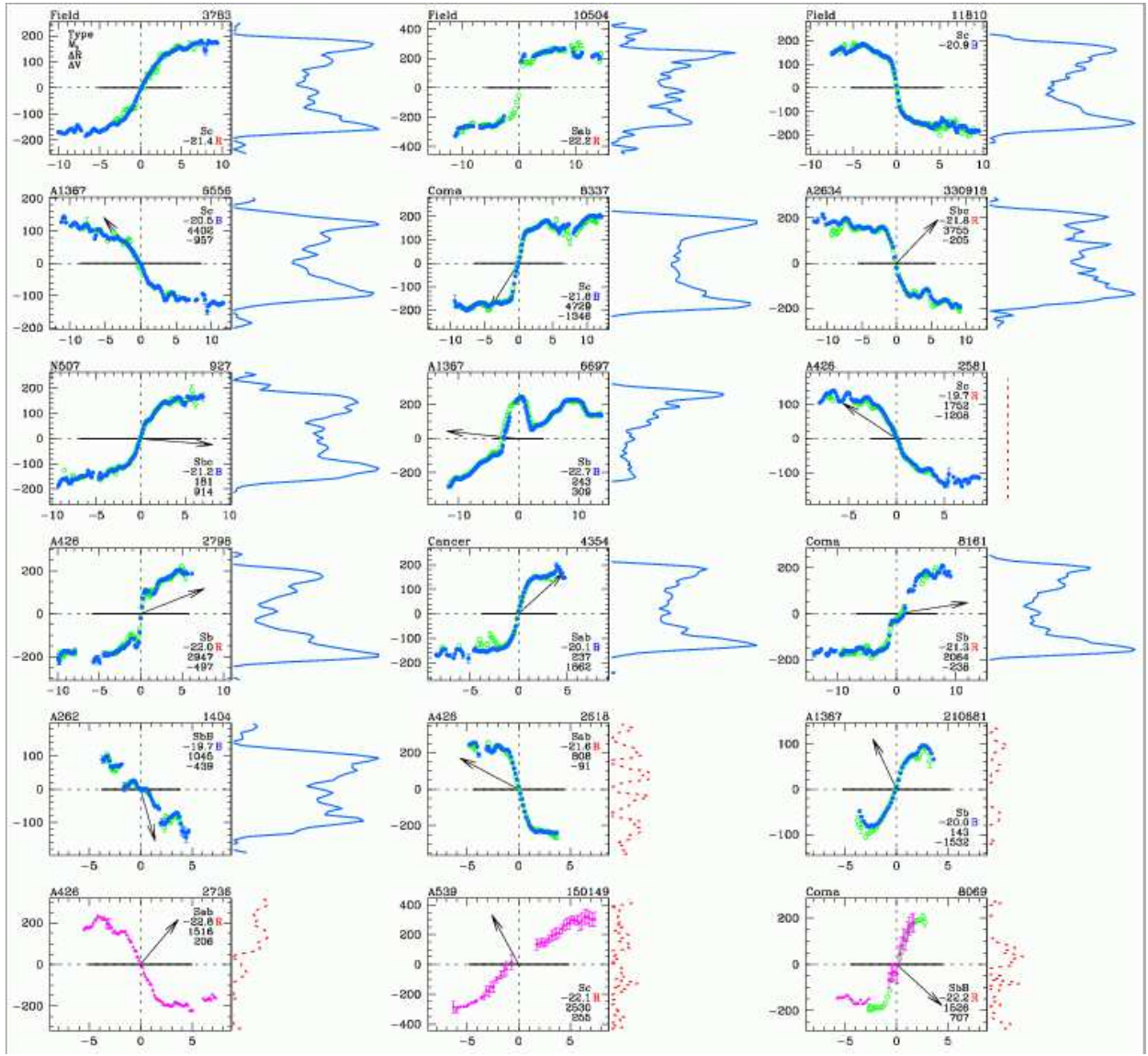,width=6.5truein}\end{center}
  \caption[RCHI sample plot]
{A representative mosaic of sets of three galaxies
which are, row by row, (1) isolated in the field, (2) more than 3 \hMpc from a
cluster, (3) within 3 \hMpc of a cluster, with {\it normal} \Halpha extent,
(4) weakly asymmetric, (5) {\it stripped}, and (6) {\it quenched}.  Boxes show
the optical rotation curve (\hkpc on the x-axis, \kms on the y-axis, centered
on the continuum) with the \HI line profile to the right (counts on the x-axis
and velocity on the y-axis, same scale as optical velocities).  \Halpha
emission flux is shown as blue solid circles, \fnii as hollow green circles,
with error bars shown only where larger than 10 \kms, and \Halpha absorption
as magenta triangles, with error bars shown only where larger than 20 \kms.
\HI flux is drawn with a blue solid line (HI normal) or a red dashed line (\HI
deficient).  The peak \HI flux is scaled to symbolize roughly the level of \HI
gas depletion; a linear \HI flux indicates that we have a measurement of the
total galaxy \HI gas mass, but its distribution in radial velocity space is not
available.  A solid $\pm2R_d$ disk length is drawn along the major axis, and
an arrow points towards the cluster center.  As our sample is made up of
fairly edge-on galaxies, a face-on encounter is suggested statistically when
the arrow is perpendicular to the disk and an edge-on one when the arrow lies
along the disk.  (Note that the angle between the arrow and the rotation curve
means nothing!)  Plots are annotated with galaxy type, $M_I$ followed by B or
R for blue or red B$-$I color, and clustercentric radius in \hkpc and velocity
offset in \kms.}
  \label{fig:RCHISample}
\end{figure*}

Figure~\ref{fig:RCHISample} shows the \Halpha, \fnii, and \HI data for a
representative subset of our data set.  By plotting our \HI line profiles on
the same velocity scale as the spatially resolved optical spectra, we can
align the frequency distribution of \HI gas to the spatial axis and estimate
its distribution along the disk.  For galaxies associated with a cluster, an
arrow points towards the cluster center.  As our sample is made up of fairly
edge-on galaxies, a face-on encounter is suggested statistically when the
arrow is perpendicular to the x-axis and an edge-on one when the arrow lies
along the x-axis.

The optical spectra of galaxies in the field or more than 3 \hMpc from the
center of clusters (first two rows) share common properties of uniform shape
and extent.  They are symmetric when centered about the continuum or the
median velocity, both in radial extent ($\sim$ 10 \hkpc), in radial strength,
and in velocity structure.  The shape of the rotation curves is moderately
smooth, characterized by a steep inner rise, an elbow turnover point, and a
relatively flat outer region.  Both \Halpha and \fnii can be traced along the
entire profile, except for the nucleus where the \Halpha may be partly
absorbed.  Large isolated \HII regions can bias the small scale flux
distribution, but the underlying structure is quite uniform. These galaxies
have the expected amounts of \HI gas, distributed in double-horned
profiles. These objects define the {\it normal}, well-behaved appearance of
optical rotation curves.

Most galaxies on the outskirts of all clusters, or located within cool (or not
particularly rich) clusters display \Halpha flux characteristics similar to
the field (\UGC927, in the third row, is representative).  Most, though not
all, contain the expected reservoir of \HI gas.  A small percentage display
weak asymmetry in the distribution of \Halpha flux (fourth row).  In these
cases, we find that the truncated \Halpha distribution is matched by a
decrease in the amount of equivalent velocity \HI gas on that side of the
disk.

In contract, galaxies with normal optical spectra are rarely found within the
cores of rich, hot clusters.  Instead, the spectra are less extended along the
disk and exhibit a greater variation in line strength, and a greater
difference between the small scale variations in velocity (\eg ripples) on the
two sides of the rotation curve.  A few, like \UGC6697 (row 3), show evidence
for large-scale distortion in the shape of the velocity profile.  The bulk of
these \Halpha spectra divide into three categories.  First, we find galaxies
for which the distribution of \Halpha flux is truncated on one side of the
disk, by at least either 5 \hkpc or 50\%, relative to the other side.  There
is good agreement between the distribution of \HI and \Halpha flux remaining
in these {\it asymmetric} galaxies, which make up the bulk of the \HI
detections within hot clusters.  Second, we find galaxies for which the
\Halpha flux extends to less than either 5 \hkpc or 3 $R_d$ across the entire
disk.  Third, we find galaxies for which there is no \Halpha emission detected
across the disk at all.  These galaxies are strongly \HI deficient, with very
few \HI detections.

In summary, we define four classes based on \Halpha emission flux properties:
(1) {\it normal}, showing properties equivalent to those found in the field,
(2) {\it asymmetric}, with \Halpha flux truncated along one side of the disk,
(3) {\it stripped}, with strong \Halpha truncation across the entire disk, and
(4) {\it quenched}, with no \Halpha emission flux.  These four terms will be
italicized through the text, to avoid confusion with that used elsewhere.  
These classes can be closely compared to previous classification of cluster
spirals (\ie van den Bergh 1999 on anemic spirals; Cayatte \etal 1990, 1994,
Guhathakurta \etal 1988 on the process of Virgo spiral infall), though we
focus on the \Halpha distribution rather than that of {\rm H\kern.2em{\smfont
I}}.  We have categorized infalling galaxies with different observables and
are thus sensitive to different markers of evolution along the infall path,
but classify the initial stages of infall (called {\it normal} here, analogous
to Group I in Cayatte \textit{et al.}) and the end-state (called {\it
quenched} here, analogous to anemic, or to Group IV for Cayatte \textit{et
al.})  similarly.

\begin{figure*} [htbp]
  \begin{center}\epsfig{file=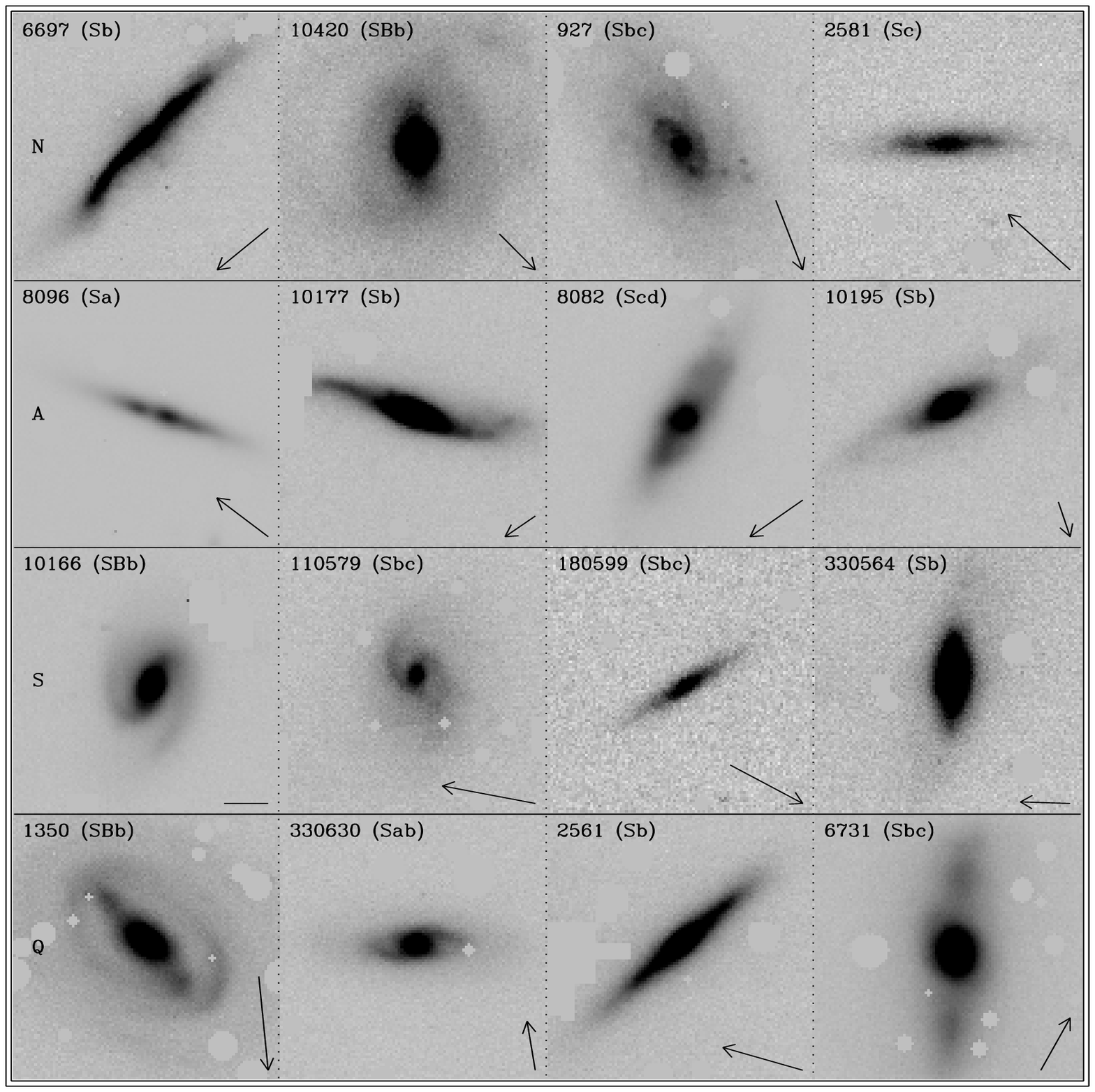,width=6.0truein,angle=-90}\end{center}
  \caption[Normal, asymmetric, stripped, and quenched galaxy images]
{A representative $I$-band mosaic of four rows of \Halpha (1) {\it normal},
(2) {\it asymmetric}, (3) {\it stripped}, and (4) {\it quenched} spirals
Postage stamps are each 1\arcmin by 1\arcmin, with north up and east to the
left, and foreground stars have been masked.  A 5 \hkpc bar shows the
direction from the galaxy to the cluster center for all galaxies within 3
\hMpc (all but \UGC10166).  Some of the spirals are fairly edge-on and thus it
can be difficult to assess the level of spiral arm structure, though the {\it
quenched} barred spiral \UGC1350 suffers from no such ambiguities.  The bulk
of the {\it normal} spirals display smooth, regular morphology, though a few
exceptions such as \UGC6697 and \UGC10420 can be found, mostly in the hot
cluster cores.  The {\it asymmetric} spirals show some distortion, consisting
of variations in the ellipticity of the central surface brightness contours,
disk contours off-center from the center of light, and some disruption in the
outer disk, but fall mainly within the range observed within the {\it normal}
spiral sample.  The {\it stripped} and {\it quenched} spirals have smooth and
regular surface brightness profiles.}
  \label{fig:NASQplate} 
\end{figure*}

Figure~\ref{fig:NASQplate} shows a representative mosaic of \Ib images of
galaxies of all four classes.  It appears that the underlying older stellar
population has not been disturbed in any of these galaxies, and the processes
which are stripping the gas and halting new star formation have not affected
the distribution of the long-lived disk population.  There are small
variations in some of the {\it asymmetric} galaxies (A), but most display a
normal \Ib morphology.  The {\it stripped} spirals (S) have a smooth
distribution; some still show spiral arm structure.  The {\it quenched} (Q)
spirals rarely have strong spiral arm structure, though note the barred spiral
\UGC1350.  Our survey selection function discriminates against galaxies of an
extremely disturbed morphology (not identifiable as spirals), in the process
of a major merger (interacting with a companion), or undergoing extreme
morphological transformation (\eg harassment).  Because of this, we are not
tracking evolutionary forces so strong as to significantly disturb the
fundamental structure of the disk.  Instead, we have focused upon galaxies
which maintain a recognizable underlying form to the disk, allowing it to
remain in place throughout the process of infall.

\begin{table*} [htbp]
  \caption{Distribution of {\rm H}\kern.2em{\smfont I} Gas Properties}
  \begin{center}
  \begin{tabular} {l r r r r r r r r r r r r r r} 
  \tableline
  \tableline
  Type & \multicolumn{5}{c}{Field Galaxies}    & \multicolumn{5}{c}{Cluster Members} & Total & 
  {\rm H\I$_{gas}$} & B/T\tablenotemark{a} & B$-$I                                      \\
  & & & & & & & & & & &
  \multicolumn{2}{r}{log(h$^2$ M$_{\sun}$)}  &               & \multicolumn{1}{c}{(mags)} \\
  \multicolumn{1}{c}{(1)}      & \multicolumn{1}{c}{(2)$^b$}  & \multicolumn{1}{c}{(3)$^c$}  & \multicolumn{1}{c}{(4)$^d$}  & \multicolumn{1}{c}{(5)$^e$}  & 
  \multicolumn{1}{c}{(6)$^f$}  & \multicolumn{1}{c}{(7)$^b$}  & \multicolumn{1}{c}{(8)$^c$}  & \multicolumn{1}{c}{(9)$^d$}  & \multicolumn{1}{c}{(10)$^e$} & 
  \multicolumn{1}{c}{(11)$^f$} & \multicolumn{1}{c}{(12)} & \multicolumn{1}{c}{(13)} & \multicolumn{1}{c}{(14)} & \multicolumn{1}{c}{(15)}  \\
  \tableline 
  \multicolumn{15}{l}{{\it Normal} (\Halpha emission consistent with isolated field)}                                                  \\
  Sa -- Sbc &  31 &   4 &   0 &   0 &   0 &  67 &  13 &   6 &   1 &  17 & 139  &  9.43(0.43)       & 0.17(0.10)       & 1.90(0.42)     \\ 
  Sc -- Sd  &  43 &   2 &   3 &   0 &   1 &  36 &   4 &   4 &   0 &   2 &  95  &  9.48(0.42)       & 0.07(0.06)       & 1.66(0.52)     \\ 
  \multicolumn{15}{l}{{\it Asymmetric} (unequal \Halpha emission from one side of disk to the other)}                                  \\
  Sa -- Sbc &   0 &   0 &   0 &   0 &   0 &   3 &   0 &   3 &   0 &   2 &   8  &  9.18(0.51)       & 0.20(0.11)       & 1.94(0.58)     \\ 
  Sc -- Sd  &   0 &   0 &   0 &   0 &   0 &   1 &   1 &   2 &   1 &   0 &   5  &  9.05(0.35)       & 0.20(0.15)       & 2.05(0.74)     \\ 
  \multicolumn{15}{l}{{\it Stripped} (truncated \Halpha emission along disk)}                                                          \\
  Sa -- Sbc &   0 &   0 &   0 &   1 &   0 &   1 &   5 &   3 &   0 &   2 &  12  &  8.64(0.38)       & 0.15(0.09)       & 2.02(0.56)     \\ 
  Sc -- Sd  &   0 &   0 &   0 &   0 &   0 &   1 &   0 &   0 &   0 &   0 &   1  &  7.86\phm{(0.05)} & 0.05\phm{(0.09)} & \nodata        \\ 
  \multicolumn{15}{l}{{\it Quenched} (\Halpha absorption along disk)}                                                                  \\
  Sa -- Sbc &   0 &   1 &   0 &   0 &   0 &   0 &   1 &  12 &   0 &   3 &  17  &  8.52(0.29)       & 0.20(0.10)       & 2.35(0.27)     \\ 
  Sc -- Sd  &   0 &   0 &   0 &   0 &   0 &   0 &   0 &   4 &   0 &   0 &   4  &  8.52(0.14)       & 0.23(0.11)       & \nodata        \\ 
  \multicolumn{15}{l}{Total}                                                                                                           \\
  Sa -- Sbc &  31 &   5 &   0 &   1 &   0 &  71 &  19 &  24 &   1 &  24 & 176                                                          \\  
  Sc -- Sd  &  43 &   2 &   3 &   0 &   1 &  38 &   5 &  10 &   1 &   2 & 105                                                          \\  
  Sa -- Sd  &  74 &   7 &   3 &   1 &   1 & 109 &  24 &  34 &   2 &  26 & 281                                                          \\  
  \tableline 
  \multicolumn{15}{l}{\hspace{0.05truein} $^a$Bulge to total fraction of \Ib luminosity} \\
  \multicolumn{15}{l}{\hspace{0.05truein} $^b${\rm H{\smfont I}-nrm}: log ${\rm H{\smfont I}_{def}} <$ log(2.5),    detected \HI normal    galaxies} \\
  \multicolumn{15}{l}{\hspace{0.05truein} $^c${\rm H{\smfont I}-def}: log ${\rm H{\smfont I}_{def}} \geq$ log(2.5), detected \HI deficient galaxies} \\
  \multicolumn{15}{l}{\hspace{0.05truein} $^d${\rm H{\smfont I}-lim}: log ${\rm H{\smfont I}_{def}} \geq$ log(2.5), from upper limit on \HI gas} \\
  \multicolumn{15}{l}{\hspace{0.05truein} $^e${\rm H{\smfont I}-nol}: upper limit on \HI gas lies within {\rm H{\smfont I}-nrm} range} \\
  \multicolumn{15}{l}{\hspace{0.05truein} $^f${\rm H{\smfont I}-non}: \HI content unknown} 
  \end{tabular}
  \end{center}
  \label{tab:h1numbers}
\end{table*}

We suggest that these four \Halpha flux classes represent successive stages of
the infall process, for relatively isolated, massive spirals interacting with
a hot gas cluster component.  Qualitatively, an incoming field spiral on the
outskirts of a cluster will have the gas reservoir and star formation
properties of a {\it normal} field spiral.  Those which encounter the
intracluster medium at a face-on orientation should be sieved of atomic gas
across the outer region of the disk simultaneously, producing an abrupt halt
to star formation due to the large cross-sectional interaction (\cf Abadi
\etal gas stripping models).  Those which enter the high density gas with a
more edge-on orientation will be exposed to strong ram pressure forces on the
leading edge of the disk, while the remainder of the gas reservoir will be
sheltered for approximately half of a full disk rotation period, assuming a
radial path of infall.  The surprising detection of a significant number of
galaxies with asymmetric distributions of \HI and \Halpha flux suggests that
the stripping process operates to quench star formation within $10^8$ years,
before the disk has rotated enough to erase directional signatures of infall.
Recently stripped galaxies will maintain star formation in the central
regions, of order 3 \hkpc in radius, where they have managed to retain a
portion of the gas (possibly also funneled to the core in more violent cases
of stripping).  Young star formation will come to a halt across the disk,
spiral structure will fade, and the galaxies will slowly drift towards the
morphologies and orbital patterns of cluster S0s.
This pattern falls broadly within the picture presented by models of infall
(\cf Abadi \etal 1999; Vollmer \etal 2001b), including simulations of
individual cases (\ie Vollmer \etal 2001a), though detailed models include
additional mechanisms to which our data are not sensitive (\eg bursts of star
formation as stripped gas falls back onto the disk, Vollmer \etal 2001b).  We
do not attempt to constrain the gas stripping process with this level of
detail, but identify a short-phase period based on observed \Halpha flux
asymmetries which suggest a shorter timescale for ram pressure stripping to
affect molecular gas and derivative star formation than is predicted by
numerical simulations.

\begin{figure*} [htbp]
  \begin{center}\epsfig{file=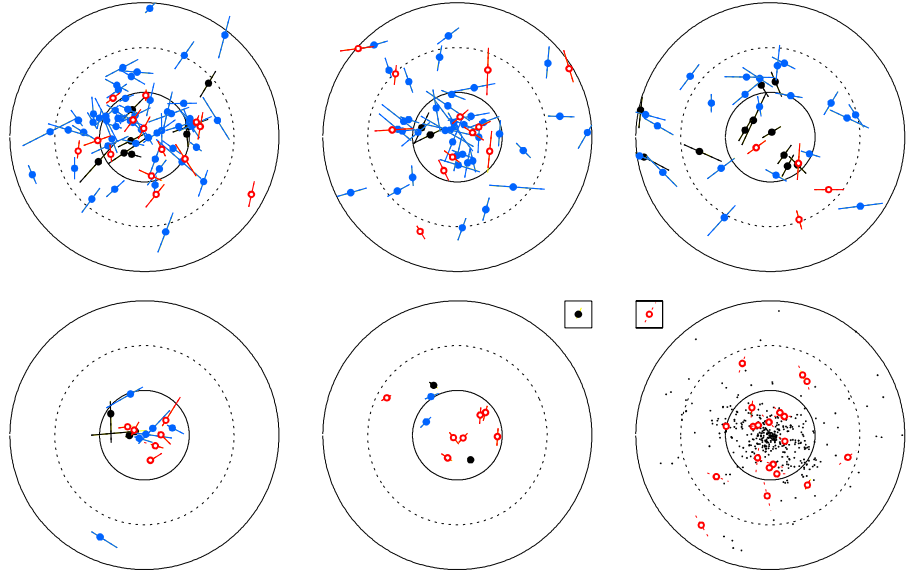,width=5.0truein}\end{center}
  \caption[Uber-cluster sample plot]
{Distribution of galaxies within 3 \hMpc of all clusters, with circles drawn
at 1, 2, and 3 \hMpc.  \HI normal galaxies are draw in blue with solid bulges,
HI deficient galaxies in red with hollow bulges, and galaxies of unknown \HI
gas mass in black with solid bulges.  Individual galaxy glyphs show the extent
of the \Halpha flux along the major axis of the disk; a solid line indicates
\Halpha emission and a dashed line (lower right panel only) indicates \Halpha
absorption.  The upper panels show {\it normal} galaxies (no striking
properties of \Halpha flux); on the right within A1656, A426, or A2199, in the
middle within the remaining hot clusters A2151, A1367, A2634, and A539 and on
the left within the cooler clusters.  The lower left/middle/right panels show
{\it asymmetric}/{\it stripped}/{\it quenched} galaxies.  Boxed galaxies in
the lower panels identify the two {\it stripped} or {\it quenched} galaxies
beyond 3 \hMpc.  Cluster S0s have been added to the lower right panel
as small black dots; their distribution peaks in the center and drops
exponentially, while there are no quenched spirals in the inner 200
\hkpc of any of the clusters.}
  \label{fig:UberSample} 
\end{figure*}

Figure~\ref{fig:UberSample} shows the spatial distribution of the spiral
galaxies, within three \hMpc of the clusters, employing symbols to distinguish
different classes and \Halpha extents.  Galaxies have been divided into six
categories, according to their \Halpha flux and the X-ray properties of their
parent clusters.  The top row contains all galaxies with {\it normal} \Halpha
properties.  From left to right, clusters are sorted into those with X-ray
temperatures cooler than 3 keV, between 3 and 4.5 keV, and hotter than 4.5
keV.  Both cool and warm clusters are characterized by a large population of
\HI normal galaxies and a smaller fraction of \HI deficient ones; the radial
galaxy distribution in all clusters peaks in the cores.  The hottest clusters
also contain galaxies with a range of \HI gas masses, but those with {\it
normal} \Halpha properties are found predominantly outside of the inner 600
\hkpc region.  In summary, \Halpha {\it normal} galaxies are found throughout
all of the clusters, but are scarce within the inner regions of the hottest
cluster cores.

The bottom row shows the spatial distribution, from left to right, of {\it
asymmetric}, {\it stripped}, and {\it quenched} spirals found in all of the
clusters.  Thirteen of the fourteen {\it asymmetric} galaxies are members of
warm (above 3 keV) clusters.  They are located preferentially in the inner 600
\hkpc region (the radial distribution differs from that of the {\it normal}
galaxies found throughout the clusters at a $>$99\% confidence level), where
the effect of ram pressure stripping from the hot gas component is expected to
become prevalent.  Both strongly HI deficient, the {\it stripped} and {\it
quenched} spirals exhibit a more relaxed radial distribution; we find them
within clusters of all temperatures.  The cores of the hottest clusters are
thus shown to be populated predominantly with {\it asymmetric}, {\it
stripped}, and {\it quenched} spirals, rather than those of the {\it normal}
class.  The remaining clusters contain galaxies with a range of \Halpha
properties in the cores, but with a high {\it normal} fraction.

Table~\ref{tab:h1numbers} summarizes the distribution of galaxies by class, HI
content, cluster membership, morphological type and \Halpha extent.  Within
each class, galaxies are divided into early and late types; mean values of the
\HI mass, bulge--to--total luminosity and \BI colors are also given.
magnitudes from the RC3 where available or else from NED.  Note that while the
derived colors are imprecise, we use them only as a secondary indicator of
star formation, binning the sample into wide color bins to identify galaxies
with extreme colors.  \HI deficiency is more prominent within clusters, as
expected, and also becomes stronger for galaxies without {\it normal} \Halpha
flux.  Both bulge fractions and \BI colors appear to increase similarly, for
{\it asymmetric} and {\it quenched} galaxies, while the situation for {\it
stripped} galaxies is more complicated, as discussed below.

\subsection{Normal Spirals}
\label{subsubsec:Normals}

We have characterized the properties of the bulk of {\it normal} spirals as
being equivalent to those found for field spirals.  However, galaxies with
{\it normal} \Halpha spectra within the inner regions ($< 900 $ \hkpc) of the
hottest clusters within our sample tend to have a smaller \Halpha extent than
expected (91\% fall below the average value), and a lower \Halpha equivalent
width.  We lack \HI measurements for a large number of these galaxies as they
lie within the cluster A2199, but those which have been observed show
extremely high \HI deficiency.

\begin{table*} [htbp]
  \caption{Mean Properties of Spiral Sub-Classes}
  \begin{center}
  \begin{tabular} {l r r r r r r r r r r r r} 
  \tableline
  \tableline
  {Population} & {$\Delta$Radius} & {$\Delta$cz} & {B/T} & {M$_I$} & {B$-$I} & 
  {R$_d$}   & \multicolumn{2}{c}{R$_b$}          & \multicolumn{2}{c}{ORC\tablenotemark{a}$_{ext}$}    & {$<$DEF$>$} & {\rm H\I$_{gas}$} \\
                       & {(\hkpc)}         & {($\sigma$)} &               & {(mag)} & {(mag)} & 
  {(kpc/h)} & \multicolumn{2}{c}{(\hkpc, R$_d$)} & \multicolumn{2}{c}{(\hkpc, R$_d$)} & \multicolumn{2}{r}{log(h$^2$ M$_{\sun}$)} \\
  \multicolumn{1}{c}{(1)}  & \multicolumn{1}{c}{(2)}  & \multicolumn{1}{c}{(3)}  & \multicolumn{1}{c}{(4)}  & \multicolumn{1}{c}{(5)}  & 
  \multicolumn{1}{c}{(6)}  & \multicolumn{1}{c}{(7)}  & \multicolumn{1}{c}{(8)}  & \multicolumn{1}{c}{(9)}  & \multicolumn{1}{c}{(10)} & 
  \multicolumn{1}{c}{(11)} & \multicolumn{1}{c}{(12)} & \multicolumn{1}{c}{(13)}  \\
  \tableline 
  {\it Normal}                     & 1194(727) & 1.0(1.4) & 0.13(0.10) & -21.5(0.9) & 1.81(0.47) & 3.0(1.2) & 13.0(4.5) & 4.7(1.3) &  9.9(3.5) & 3.6(1.2) &  0.09(0.33) & 9.45(0.43)  \\ 
  ~~Field                          & \nodata   & \nodata  & 0.10(0.09) & -21.2(0.7) & 1.74(0.44) & 2.9(1.0) & 13.6(2.9) & 5.0(1.4) & 10.8(3.1) & 4.0(1.3) &  0.00(0.23) & 9.60(0.26)  \\ 
  ~~Clusters                       & 1194(727) & 1.0(1.4) & 0.14(0.10) & -21.5(0.9) & 1.82(0.48) & 3.0(1.3) & 12.9(4.7) & 4.6(1.3) &  9.8(3.5) & 3.6(1.2) &  0.10(0.33) & 9.42(0.43)  \\ 
  ~~Early                          & 1167(759) & 1.1(1.2) & 0.17(0.10) & -21.7(0.8) & 1.90(0.42) & 3.0(1.3) & 13.1(4.5) & 4.6(1.2) &  9.8(3.7) & 3.5(1.1) &  0.10(0.32) & 9.43(0.41)  \\ 
  ~~Late                           & 1357(681) & 2.0(2.7) & 0.07(0.06) & -21.1(0.8) & 1.67(0.52) & 2.9(1.1) & 12.9(4.6) & 4.8(1.4) & 10.2(3.3) & 3.8(1.3) &  0.07(0.32) & 9.48(0.42)  \\ 
  ~~{\rm H\I-nrm}\tablenotemark{b} & 1265(728) & 1.2(1.3) & 0.13(0.10) & -21.5(0.9) & 1.77(0.49) & 3.0(1.2) & 13.2(4.3) & 4.7(1.3) & 10.4(3.5) & 3.7(1.2) & -0.03(0.20) & 9.58(0.30)  \\ 
  ~~{\rm H\I-def}\tablenotemark{c} &  951(486) & 1.1(1.2) & 0.15(0.11) & -21.4(1.0) & 1.91(0.36) & 2.6(1.1) & 11.6(4.3) & 4.9(1.2) &  6.9(2.6) & 3.1(1.2) &  0.55(0.13) & 8.91(0.25)  \\ 
  ~~{\rm H\I-lim}\tablenotemark{d} & 1612(909) & 3.3(4.2) & 0.12(0.10) & -21.3(0.6) & 1.84(0.32) & 2.6(0.9) & 12.5(4.2) & 4.8(1.2) &  8.5(2.5) & 3.4(1.2) &  0.90(0.25) & 8.58(0.34)  \\ 
  \multicolumn{1}{l}{~~Hot cores\tablenotemark{e}}
                                   &  557(233) & 1.1(0.6) & 0.12(0.05) & -21.8(0.8) & 2.01(0.36) & 3.0(1.1) & 12.3(5.0) & 4.4(0.7) &  8.0(3.7) & 2.9(0.7) &  0.29(0.30) & 9.26(0.21)  \\ 
  {\it Asymmetric}                 &  587(607) & 1.0(0.7) & 0.19(0.11) & -21.8(0.8) & 1.97(0.58) & 3.2(1.7) & 13.5(5.4) & 4.5(1.5) & 10.5(6.0) & 3.7(1.8) &  0.41(0.46) & 9.18(0.48)  \\ 
  ~~B$-$I $\le$ 1.5                &  473(178) & 0.8(0.6) & 0.09(0.06) & -21.2(0.8) & 1.29(0.17) & 2.1(1.1) & 10.4(4.1) & 5.3(1.7) &  8.3(4.9) & 4.5(2.4) &  0.64(0.26) & 8.84(0.27)  \\ 
  ~~B$-$I $\ge$ 2.0                &  715(787) & 1.0(0.7) & 0.25(0.09) & -22.1(0.7) & 2.31(0.35) & 3.6(1.9) & 13.6(5.0) & 4.2(1.5) & 10.2(4.9) & 3.1(1.5) &  0.25(0.47) & 9.34(0.54)  \\ 
  {\it Stripped}                   &  728(490) & 1.6(1.5) & 0.14(0.09) & -20.7(1.1) & 2.02(0.56) & 2.1(0.8) &  9.2(3.1) & 4.4(0.9) &  4.1(0.9) & 2.1(0.7) &  0.74(0.37) & 8.58(0.43)  \\ 
  {\it Quenched}                   & 1019(609) & 0.8(0.6) & 0.20(0.10) & -21.9(0.5) & 2.33(0.25) & 2.3(0.5) & 10.5(2.8) & 4.8(1.4) &  5.4(2.0) & 2.7(0.7) &  0.95(0.28) & 8.52(0.20)  \\ 
  \tableline 
  \multicolumn{13}{l}{\hspace{0.05truein} $^a$The maximum extent of H$\alpha$, or of {\rm [N\II]} emission.} \\
  \multicolumn{13}{l}{\hspace{0.05truein} $^b$log ${\rm H\I_{def}}    <$ log(2.5).} \\
  \multicolumn{13}{l}{\hspace{0.05truein} $^c$log ${\rm H\I_{def}} \geq$ log(2.5).} \\
  \multicolumn{13}{l}{\hspace{0.05truein} $^d$log ${\rm H\I_{def}} \geq$ log(2.5), from upper limit on \HI gas.} \\
  \multicolumn{13}{l}{\hspace{0.05truein} $^e$Galaxies within clusters A1656, A426, and A2199, for which kT $>$ 4 keV;  within 900 \hkpc of the cores.} \\
  \end{tabular}
  \end{center}
  \label{tab:NASQ}
\end{table*}

In the inner 600 \hkpc region, where the hot X-ray gas component becomes
significant and ram pressure stripping a serious concern, there are only five
\Halpha {\it normal} galaxies found within A1656, A426, and A2199.  Three lie
on the 600 \hkpc outskirts and are oriented within 10$^{\circ}$\ of a face-on
trajectory relative to the cluster cores, suggesting a face-on infall path for
a radial orbit.  The remaining two are oriented more edge-on and have a
maximum 6 \hkpc \Halpha extent, on the edge of our 5 \hkpc {\it stripped}
criterion.  These galaxies fit a picture in which no spiral passes through a
hot core without substantial alteration.

Most galaxies within the cooler (kT $<$ 4 keV) clusters show far less evidence
for substantial disruption.  Many galaxies within the inner regions of A1367,
A2151, A539, and A2634 display \Halpha {\it normal} spectra, with
characteristics similar to those of the field.  Those of moderately truncated
\Halpha extent (25\%) tend to be strongly \HI deficient.  There are seven
galaxies, for example, with {\it normal} \Halpha spectra within 600 \hkpc of
the core of A1367 (close neighbor to A1656), most with quite normal
properties.  We take special note, however, of \UGC6697, which has an extended
but very disturbed rotation curve, characterized by very broad and strong
\Halpha emission extending far beyond the nucleus.  This well-studied object
(Sullivan \etal 1981; Kennicutt, Bothun, \& Schommer 1984, and references
therein; Gavazzi \etal 2001) has a very high rate of star formation
(EW$_{H\alpha}$ = 61\AA), and the \Ib image shows extreme distortion along the
large edge-on disk, flaring at both ends.  Nulson (1982) argued that the
primary gas removal mechanism is turbulent viscous stripping rather than ram
pressure, which would be inadequate to produce so strong an effect.  More
recently, Gavazzi \etal have combined narrowband \Halpha and broadband optical
images with longslit observations taken at a variety of positions and position
angles and Fabry-Perot interferometry and conclude that the object is composed
of two interacting galaxies, the data for which are complicated by the
presence of a superimposed galaxy lying directly in the background.  Due to
the presence of the background galaxy in the optical spectra, this object was
not included when the {\it normal} sample was characterized (see
Tables~\ref{tab:h1numbers} and ~\ref{tab:NASQ}).

\subsection{Asymmetric Spirals}
\label{subsubsec:Asymmetrics}

It is well known that lopsidedness is a common feature in both the optical and
HI disks of seemingly undisturbed galaxies (e.g., (Richter \& Sancisi 1994;
Rix \& Zaritsky 1995; Kornreich, Haynes, Lovelace, \& van Zee 2000). Weak
trends in \Halpha extent and asymmetry with clustercentric radius were found
for \Halpha strong detections in the nearby universe (Dale \etal 1999), though
not for a similar sample of redshift $z \sim 0.1$ clusters (Dale \& Uson
2003).  The Virgo spiral \NGC4522 (Kenney \& Koopmann 1999) is a clear case of
a peculiar \Halpha flux distribution associated with ram pressure stripping
across the disk of a galaxy.  Indeed, moderate asymmetry, both dynamical and
morphological, is common, occurring in 30\% to 50\% of galaxy disks located in
a broad range of environments.

Here we use simple but robust criteria to distinguish the {\it asymmetric}
spirals by quantifying the degree and extent of asymmetry observed in their
\Halpha rotation curves.  The determination of the extent is relatively
insensitive to the specific detection criteria used when tracing the spectra,
because the drop in signal to noise is quite abrupt (the spectra tend to
terminate as a step function rather than gradually tapering off in intensity).
A truncation in the radial extent of star formation is directly connected to
an extreme curtailment of the gas reservoir.  In contrast, the equivalent
width of the \Halpha line flux is sensitive both to galaxy type and to
extinction, and can be enhanced by some interactions (\eg \UGC6697, where gas
funneled to the core is stimulating a starburst phase) or diminished by
others.

\begin{figure} [htbp]
  \begin{center}\epsfig{file=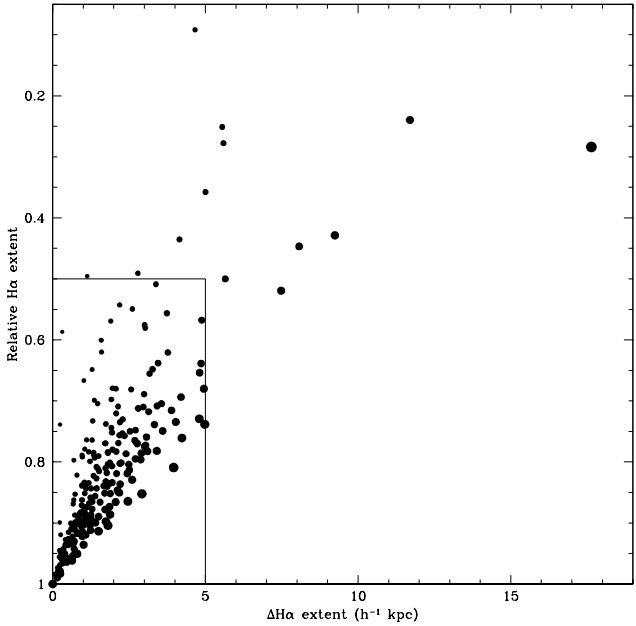,width=3.4truein}\end{center}
  \caption[asym indices plot]
{The distribution of \Halpha flux asymmetry indices throughout the sample.
The x-axis shows the differential extent of \Halpha (longest -- shortest radial
extent, from one side of the disk to the other), and the y-axis the relative
\Halpha extent (shortest over longest extent).  Point size scales slightly 
with total \Halpha extent, and the most extended galaxies lie along the lower
right-hand edge of the point distribution.  The {\it asymmetric} galaxies fall
outside of the locus in which most galaxies are found, bounded at a
differential extent of 5 \hkpc and a relative extent of 50\%.  Most {\it
asymmetric} galaxies fall beyond both limits, in the upper right-hand corner
of the plot, and all have a relative extent less than 55\%.}
  \label{fig:2asms}
\end{figure}

In order to avoid asymmetry caused by a single isolated HII region or other
peculiarity, we settled on two specific criteria for our {\it asymmetric}
classification: the radial extents of \Halpha emission traced on each side of
the disk must either differ by more than 5 \hkpc ($\Delta$ \Halpha $>$ 5\hkpc)
or form a ratio ($r_1/r_2$) of less than 1:2.  The percentage of galaxies
which met these criteria was less than 5\% and all but one fall within 1 \hMpc
of a cluster core; the adopted criteria do indeed identify the extreme cases
of asymmetry.  
arbitrary environments.  Most of the 14 {\it asymmetric} spirals show a
difference in \Halpha extent of less than 1:2, and for half of them this
difference is also more than 5 \hkpc.  In several cases the remaining \Halpha
flux on the truncated side (within the radius of truncation) is less strong
than the flux on the other side of the nucleus at the same radius, but this is
not a requirement.  Figure~\ref{fig:2asms} shows the distribution of the
dynamic sample in terms of the two key asymmetry parameters.  The {\it
asymmetric} galaxies are clear outliers from the locus of the general
distribution, along both parameter axes.

As discussed in \pone, rotation curve centerpoints were determined by either
balancing the two sides of the profile to determine a kinematic centerpoint,
or by the spatial position of the center-of-light (COL) of the continuum.  The
{\it asymmetric} galaxies are preferentially COL-centered galaxies (four out
of fourteen, or 29\%, versus 12\% for the complete sample).  This is expected,
as the kinematic measurement can be biased strongly when one side of the
rotation curve is truncated, while the COL will remain unchanged.  As an
additional check, we folded the rotation curves about the centerpoints to
verify that the inner profiles agreed on both sides of the nucleus.
Figure~\ref{fig:UberInfall} allows this check to be performed visually for the
{\it asymmetric} sample; the selected centerpoints produce an acceptable
match.

\begin{figure*} [htbp]
  \begin{center}\epsfig{file=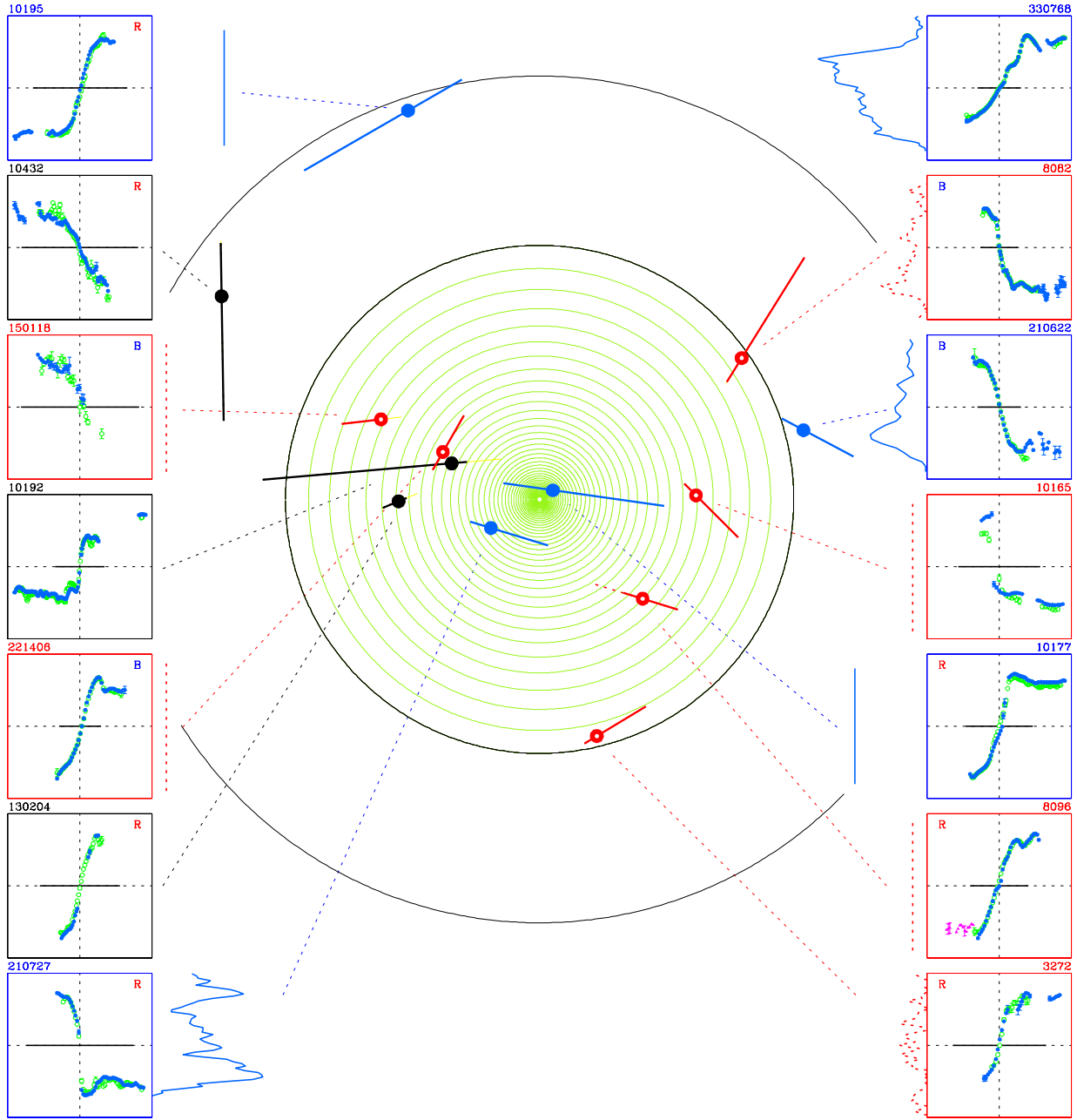,width=6.5truein}\end{center}
  \caption[Uber-cluster asym plot]
{The spatial distribution of galaxies with {\it asymmetric}
\Halpha flux extent.  The outer circle is drawn at 1 \hMpc and green mock hot
X-ray gas contours fill the inner 600 \hkpc.  Individual galaxy glyphs show
the H$\alpha$ flux along the major axis of the disk; a solid line indicates
\Halpha emission, and a dotted line \Halpha absorption (\UGC8096/\IC3949 only).  
For \UGC10192 and \AGC130204/\CGCG540-115, the \fnii\ flux (yellow) extends beyond the
\Halpha.  Galaxies with normal amounts of \HI are drawn in blue, while \HI 
deficient galaxies are drawn in red (black for no \HI data).  Optical 
rotation curves and \HI line profiles are shown as defined in 
Figure~\ref{fig:RCHISample}, and each galaxy is identified by its \AGC 
number.  The disks tend to point towards the cluster centers, suggesting 
edge-on rather than face-on infall into the hot gas.  \AGC330768/\CGCG476-112 (upper 
right) is the sole {\it asymmetric} spiral found beyond 1 \hMpc, within the 
envelope surrounding A2634.  The disk of this five-armed, \HI rich spiral is 
clearly disturbed, with a gigantic \HII region at 5 \hkpc and a perturbing 
companion galaxy well within 20 \hkpc.}
  \label{fig:UberInfall} 
\end{figure*}

We note that while there are cases where \fnii\ but not \Halpha could be
traced successfully through the nuclear region, there are few cases, primarily
{\it asymmetric}, in which the \fnii\ flux extended to larger radius than the
\Halpha flux.  This is true for cases of both \Halpha emission, typically
associated with strong \fnii, and \Halpha absorption rotation curves, where
\fnii\ can be difficult to detect at all at any radius.  We elected to measure
asymmetry purely from the \Halpha distributions, as the \fnii\ flux strength
did not necessarily correlate with the \Halpha emission strength, and the
increase in overall signal strength caused by the addition of the \fnii\ data
would have been counterbalanced by a blurring of the most interesting spectra,
where \Halpha flux was severely truncated.

Many spiral galaxies spread throughout the dynamical sample show a smaller
difference in \Halpha emission extent, between 3 and 5 \hkpc.  When we relaxed
the 5 \hkpc criteria slightly, a large number of the additional, moderately
asymmetric galaxies were found, many within the Cancer cluster.  This suggests
that less extreme forms of asymmetry may be a signature of loose group
interactions, where the low relative velocities enable galaxy-galaxy
interactions at an elevated rate.

We added a single galaxy (\UGC8096/\IC3949) to the {\it asymmetric} sample due
to highly asymmetric \Halpha absorption flux.  This reddened, \HI deficient
galaxy in the core of the Coma cluster exhibits \Halpha emission in the
nuclear region and along one side of the disk, while the edge of the other
side of the disk, pointing inwards towards the cluster core, shows only strong
\Halpha absorption.  \UGC8096/\IC3949 is the sole example of this pattern of
extreme asymmetric \Halpha absorption within the dynamical sample.

Figure \ref{fig:UberInfall} illustrates the salient properties of these {\it
asymmetric} galaxies.  They have been drawn together into a single cluster
diagram, offset from the centerpoint with the spatial offsets that they each
have relative to their individual cluster centers.  The contoured inner region
extends from the cluster core to 600 \hkpc, an upper limit on where an
infalling spiral could begin interacting significantly with the intracluster
medium in a rich cluster.  The complete dynamical sample extends to well
beyond 2 \hMpc for most of the sampled clusters, so these galaxies are clearly
located preferentially in the cores.

\AGC330768/\CGCG476-112 is the sole {\it asymmetric} spiral found beyond 1
\hMpc from a cluster core, located at a radius of 2.5 \hMpc on the outskirts
of A2634.  This galaxy differs from the other {\it asymmetric} galaxies in
several important respects: it is interacting with a companion which lies
within 16 \hkpc, it is extremely gas-rich, with an \HI deficiency measurement
of -0.44, and its \BI color places it in the center of the distribution for
{\it normal} spirals.  The cause of the asymmetry is clearly the current,
large-scale interaction with a near neighbor; the \Ib image shows a disturbed
disk, with five arms and a gigantic \HII region at a radius of 10 \hkpc which
outshines even the nuclear continuum.  The asymmetry is not caused directly by
A2634, though the frequency of near neighbors is indirectly enhanced by the
overdensity of galaxies, and groups of infalling galaxies, around the cluster.

The individual galaxy glyphs in Figure \ref{fig:UberInfall} have been drawn
with a straight bar representing the \Ib disk as if edge-on, where the bar
length shows the extent of the disk in kiloparsecs and the relative length of
the two sides shows the differing extent of the \Halpha flux on each side of
the galaxy.  We observe that the truncated sides of the disks tend to point
inwards, along the direction that one would expect the galaxy to travel on a
first infall path into the cluster core on a predominantly radial orbit.
Observational evidence indicates that infalling \HI deficient spirals tend to
lie on radial orbits (Dressler 1986), and recent simulations of cluster
dynamics (Moore \etal 1999) also find a preference for radial orbits within
the spiral population, in support of such a pattern.  For 8 of the 13 galaxies
within the cores (we exclude outlier \AGC330768/\CGCG476-112) the truncated
disk points toward the cluster core rather than away; this ratio rises to 8
out of 11 galaxies in the inner 600 \hkpc region.  The median angle formed by
the truncated side of the galaxy disks and the vectors pointing towards the
center of the cluster is 43$^{\circ}$, half of the 86$^{\circ}$\ found for the
{\it normal} spiral sample.  This angle should lie near to 90$^{\circ}$\ for a
randomly oriented sample, and the probability of finding a mean value less
than 45$^{\circ}$\ is less than 0.2\%.

Of the galaxies pointing away from the cluster centers, \UGC10195 and
\UGC10432 are on the outskirts of the cores of warm clusters (radius $\sim$
930 \hkpc), and \UGC10195 has both a normal \HI gas mass and an apparent
neighbor at 70 \hkpc.  \UGC3272 and \AGC221406/\IC4040, with major axes almost
perpendicular to the clustercentric vectors, have been considerably stripped
of \HI gas and may have recently passed through cluster centers to appear on
the other side, and \AGC210727/\CGCG097-125 is a gas-rich spiral quite near to
the center of A1367 (120 \hkpc, with a velocity offset of $+2\sigma$) which
may be infalling into the cluster from the foreground.  Without knowledge of
the specific orbits of all of the {\it asymmetric} galaxies, we cannot claim
that they show direct evidence for truncation on the leading edge of the
disks, but the probability that the correlation between position and
clustercentric angles is random is small.

Figure~\ref{fig:UberInfall} also shows the \Halpha and \HI flux for each
galaxy.  The \Halpha spectra have been centered within the display boxes from
left to right, so that the continuum emission would lie at the center of each
box if shown.  A gap on one side of the box thus indicates that the \Halpha
flux does not extend as far along that side of the disk as on the other side.
Half of the truncated spirals are extremely \HI deficient, while the remainder
range from a factor of two in \HI deficiency down to normal gas content.  The
average \HI deficiency is greater than that of the normal spiral population by
a factor of 1.75.  \HI fluxes have been plotted on the same velocity scale as
the \Halpha rotation curves to demonstrate the relative amount of \HI gas at
each point along the disk, for detected galaxies.  We observe that the
truncated sides tend to show diminished \HI flux, even within those spirals
which still contain a normal amount of \HI gas, in agreement with the trend
for the weakly asymmetric galaxies as shown in Figure~\ref{fig:RCHISample}.

The asymmetric spirals are found predominantly within X-ray warm (kT $>$ 4
keV) clusters.  Table~\ref{tab:asym} lists some of their key characteristics,
in order of decreasing cluster X-ray temperature.  Columns [14] and [15]
contain the differential and relative extents of \Halpha emission flux, our
quantified measures of asymmetry.  Several exhibit extremely strong, broad
nuclear emission, suggesting that a recent infusion of gas into the galaxy
core may have stimulated a starburst phase.  The galaxies have a bimodal
distribution in \BI above and below the mean color of the normal spirals
(1.86), peaking at 2.43 (red, near to the color of the quenched spirals at
2.28) and at 1.42 (blue).  These are the colors one would expect for a normal
cluster spiral first entering a stimulated starburst phase (SB) and then
reddening by $\sim$ 1 magnitude in a quiescent post-starburst phase (PSB).

\begin{table*} [htbp]
  \caption{Properties of Asymmetric Spirals}
  \begin{center}
  \begin{tabular} {l l r r l r r c r r r r r r r r r r r r} 
  \tableline
  \tableline
  {Cluster} & \multicolumn{2}{c}{Name \hskip0.60in$\Delta$Radius\tablenotemark{a}} & {$\Delta$cz}  & {Type} & {B/T} & {M$_I$}   & {B$-$I} & 
  {R$_d$}   & {R$_b$} & {R$_b$}   & \multicolumn{6}{c}{Optical Rotation Curve Extent\tablenotemark{b}} & {$<$DEF$>$} & {\rm H\I$_{gas}$} 
                    & {$\Delta \Theta^c$} \\
                    &  &                                                       & {($\sigma$)}  & &                              & {(mag)}   & {(mag)} & 
                    &                 & {(R$_d$)} & & & & & {(R$_d$)} & {(R$_d$)}                      & \multicolumn{2}{r}{log(h$^2$ M$_{\sun}$)}
   
                    & {($^{\circ}$)} \\
  \multicolumn{1}{c}{(1)}  & \multicolumn{1}{c}{(2)}  & \multicolumn{1}{c}{(3)}  & \multicolumn{1}{c}{(4)}  & \multicolumn{1}{c}{(5)}  & 
  \multicolumn{1}{c}{(6)}  & \multicolumn{1}{c}{(7)}  & \multicolumn{1}{c}{(8)}  & \multicolumn{1}{c}{(9)}  & \multicolumn{1}{c}{(10)} & 
  \multicolumn{1}{c}{(11)} & \multicolumn{1}{c}{(12)} & \multicolumn{1}{c}{(13)} & \multicolumn{1}{c}{(14)} & \multicolumn{1}{c}{(15)} & \multicolumn{1}{c}{(16)} & 
  \multicolumn{1}{c}{(17)} & \multicolumn{1}{c}{(18)} & \multicolumn{1}{c}{(19)} & \multicolumn{1}{c}{(20)}  \\
  \tableline 
  A1656  & \AGC221406/\IC4040\tablenotemark{d} &  248 &  0.83 & Scd  &  0.07   & -20.77  & 1.38    &  1.3    &  6.3 &  4.9    &  2.7  &  5.5 &  2.8  & 0.49 &  2.11   &  4.33   & $\ge$ 0.51 & $\le$ 8.64 &  97 \\ 
  A1656  & \UGC8082/\NGC4848           &  582 &  0.27 & Scd  &  0.06   & -22.09  & 1.50    &  2.2    & 15.7 &  7.1    &  3.7  & 15.4 & 11.7  & 0.24 &  1.75   &  6.93   &       0.78 &       9.02 &  23 \\ 
  A1656  & \UGC8096/\IC3949\tablenotemark{e}   &  305 &  0.58 & Sa   &  0.25   & -21.80  & 2.10    &  1.8    & 10.5 &  5.9    &  2.6  &  4.8 &  2.2  & 0.54 &  2.70   &  3.62   & $\ge$ 0.71 & $\le$ 8.65 &  20 \\ 
  A426   & \AGC130204/Z540-115         &  333 &  1.34 & Sc   &  0.31   & -21.67  & 2.91    &  2.5    &  7.7 &  3.2    &  1.1  &  2.2 &  1.1  & 0.49 &  0.96   &  1.09   & $\ge$ 0.17 & $\le$ 9.27 &  21 \\  
  A2151  & \UGC10192\tablenotemark{f}  &  230 & -1.12 & Sb   & \nodata & -22.33  & \nodata &  4.6    & 14.5 &  3.2    &  7.0  & 24.6 & 17.6  & 0.28 &  1.51   &  5.33   & \nodata    &    \nodata &  31 \\  
  A2151  & \UGC10165                   &  369 &  2.19 & ScdB & \nodata & \nodata & \nodata & \nodata & 24.2 & \nodata &  2.2  &  7.7 &  5.6  & 0.28 & \nodata & \nodata & $\ge$ 0.79 & $\le$ 9.40 &  47 \\ 
  A2151  & \UGC10195                   &  970 & -0.40 & Sb   &  0.35   & -22.47  & 2.02    &  5.6    & 19.4 &  3.5    &  8.1  & 15.5 &  7.5  & 0.52 &  1.44   &  2.78   &       0.01 &       9.78 & 101 \\ 
  A2151  & \UGC10177                   &   38 & -1.41 & Sb   &  0.16   & -23.42  & 2.10    &  3.7    & 17.8 &  4.8    &  6.5  & 14.6 &  8.1  & 0.45 &  1.74   &  3.90   &      -0.06 &       9.77 &  42 \\ 
  A1367  & \AGC210622/Z097-062         &  644 &  1.75 & Sb   &  0.04   & -20.25  & 1.15    &  1.3    &  8.3 &  6.3    &  3.2  &  7.4 &  4.2  & 0.44 &  2.40   &  5.52   &       0.19 &       9.26 &  43 \\ 
  A1367  & \AGC210727/Z097-125         &  133 &  2.27 & Sc   &  0.34   & -20.95  & 2.43    &  3.3    &  7.8 &  2.4    &  2.8  &  7.8 &  5.0  & 0.36 &  0.85   &  2.39   &      -0.01 &       9.43 & 133 \\ 
  A539   & \AGC150118                  &  419 & -0.51 & Sb   &  0.18   & -21.73  & 1.15    &  3.6    & 11.4 &  3.2    &  0.5  &  5.1 &  4.6  & 0.09 &  0.74   &  1.42   & $\ge$ 0.70 & $\le$ 8.80 &  34 \\ 
  A539   & \UGC3272                    &  575 &  0.51 & Sb   &  0.11   & -22.04  & 2.51    &  2.8    & 14.0 &  4.9    &  1.9  &  7.1 &  5.6  & 0.25 &  0.65   &  2.60   & $\ge$ 0.85 & $\le$ 8.80 & 107 \\ 
  A2197  & \UGC10432                   &  891 &  1.06 & Sb   &  0.31   & -22.09  & 2.57    &  7.2    & 20.0 &  2.8    &  6.9  & 16.2 &  9.2  & 0.43 &  1.00   &  2.25   & \nodata    &    \nodata & 124 \\ 
  A2634  & \AGC330768/Z476-112\tablenotemark{g} & 2478 &  0.24 & Sb   &  0.15   & -22.18  & 1.84    &  1.9    & 11.6 &  6.3    &  5.7  & 11.3 &  5.7  & 0.50 &  3.04   &  6.08   &      -0.44 &       9.98 &  82 \\ 
  \tableline 
  \multicolumn{20}{l}{\hspace{0.05truein} $^a$All length scales are expressed in units of \hkpc, unless explicitly stated otherwise.} \\
  \multicolumn{20}{l}{\hspace{0.05truein} $^b$The minimum, maximum, and differential extent of H$\alpha$ or {\rm [N\II]} emission followed by the ratio, and the minimum and maximum in units of R$_d$.} \\
  \multicolumn{20}{l}{Columns 14 and 15 contain the parameters used to define the asymmetry index.} \\
  \multicolumn{20}{l}{\hspace{0.05truein} $^c$Angle between truncated side of disk and direction to cluster centerpoint.} \\
  \multicolumn{20}{l}{\hspace{0.05truein} $^d$Bravo-Alfaro \etal (2000) report a detection in {\rm H\I}, with HI$_{def}$ = 0.61, HI$_{gas}$ = 8.52.} \\
  \multicolumn{20}{l}{\hspace{0.05truein} $^e$Bravo-Alfaro \etal (2000) report a deeper observation in {\rm H\I}, with HI$_{def}$ $\ge$ 1.9, HI$_{gas}$ $\le$ 7.5.
                                           Exhibits highly asymmetric \Halpha absorption flux } \\
  \multicolumn{20}{l}{(see Figure~\ref{fig:UberInfall}).} \\
  \multicolumn{20}{l}{\hspace{0.05truein} $^f$Huchtmeier \& Richter (1989) report a detection in {\rm H\I}.} \\
  \multicolumn{20}{l}{\hspace{0.05truein} $^g$A five armed spiral, interacting with companion at 16 \hkpc.  Note large \HI gas mass.} \\
  \end{tabular}
  \end{center}
  \label{tab:asym}
\end{table*}

The correlation between \Halpha maximum extent and truncation, and \HI gas
deficiency, is shown clearly in Figure~\ref{fig:RstrRd}.  Early type
asymmetric galaxies which still contain the bulk of the initial \HI gas
reservoirs are shifting down within the diagram towards the edge of the
envelope in which \HI normal spirals are found, though the maximum extent of
\Halpha places them well within the normal distribution.  Their \HI deficient
counterpoints have a maximum \Halpha extent which places them already at the
edge of the envelope, and star formation is being suppressed across the
remaining inner portion of the disk.  The situation is more complicated for
the few {\it asymmetric} late type spirals, which show extreme spatial
truncation of star formation regardless of \HI gas mass.

\begin{figure*} [htbp]
  \begin{center}\epsfig{file=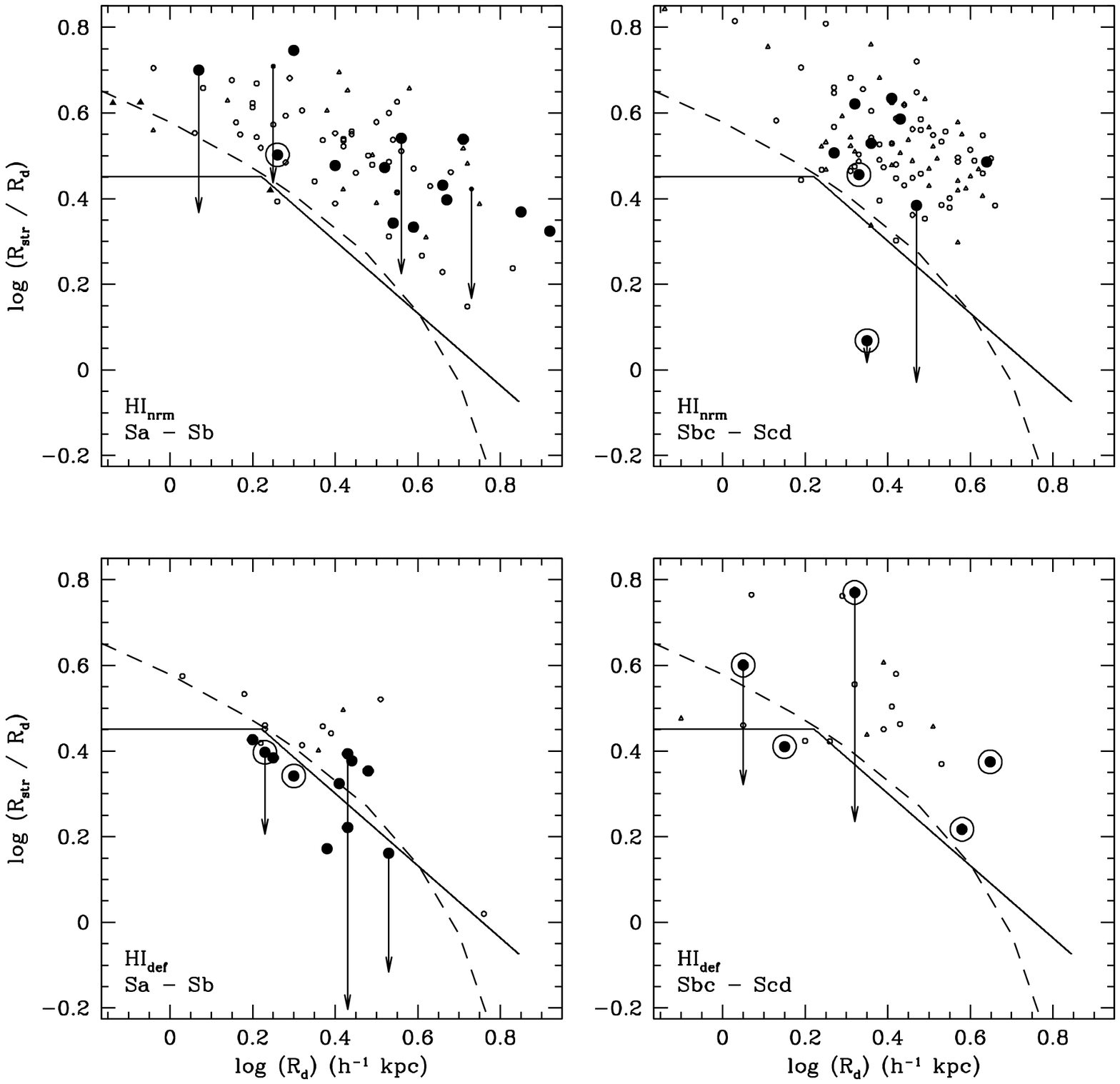,width=5.0truein}\end{center}
  \caption[RAM pressure stripping plot]
{The radial stripping radius, the maximum extent of \Halpha and \fnii
emission, as a function of disk scale length, divided into early {\bf (left)}
and late {\bf (right)} types, and \HI normal {\bf (top)} and \HI deficient
{\bf (bottom)} spirals.  Maximum and minimum extents of \Halpha are connected
with an arrow for {\it asymmetric} spirals.  Small open triangles represent
isolated field galaxies, and small open circles galaxies within cool clusters
or located more than 900 \hkpc from the core of a warm (kT $>$ 4 keV) cluster.
Core members of warm clusters are shown with a large solid circle; those few
within the cores of the three hottest clusters (A1656, A426, and A2199) are
encircled.  We have added as solid triangles in the upper left plot three
early type Virgo spirals undergoing stripping taken from Cayatte \etal (1994),
where the stripping radius has been determined from two-dimensional maps of
the \HI extent.  The solid lines define our criterion for {\it stripped}
spirals, and the dashed line a model stripping radius from Abadi
\etal (1999).}
 \label{fig:RstrRd} 
\end{figure*}

\UGC3272 and \AGC150118, like most of the spirals in the inner 1.5 \hMpc of
A539, are offset only a few hundred \kms from the cluster centroid velocity
and fall well within the cluster dispersion (see Figure \ref{fig:mship3}).
The nuclear \Halpha emission for \UGC3272 is broad and strong, while
\AGC150118 has extremely faint \Halpha and \fnii 6584\AA; both are undetected
in \HI.

The remaining galaxies are divided between the clusters with fairly high X-ray
luminosities.  Two galaxies are drawn from the core of A1367.
\AGC210727/\CGCG097-125 has broad and strong nuclear \Halpha emission, normal
\HI gas content, reddish color, and with a velocity offset of 2$\sigma$ may be
infalling from the foreground.  \AGC210622/\CGCG097-062 is \HI deficient by a
factor of two, the distribution of the remaining \HI gas clearly follows the
spatial bias of the truncated \Halpha disk.  The \Ib image hints that the disk
may be fainter on the truncated side, while more extended on the other side.
Three more galaxies are found within A2151.  The \HI profile for \UGC10177 has
been published previously (Figure 3 of Giovanelli \& Haynes 1985).  It shows
more \HI flux on the higher velocity side, in agreement with the optical
rotation curve.  The optical velocity map of Amram \etal (1992) further
confirms the observed asymmetry in the \Halpha distribution.  The \Ib image
suggests an increase in flux in front of the nucleus (as does that of
\UGC8096/\IC3949).  \UGC10165 is an extremely {\it asymmetric}, \HI deficient
galaxy in the core region, at a radius of 370 \hkpc.  \UGC10192 has broad and
extremely strong nuclear \Halpha emission.  We have no \HI data for the
galaxy, but the strong, extended \Halpha emission suggests that the gas may
still be present.

The last four galaxies are all \HI deficient by more than a factor of six, and
all lie within 600 \hkpc of the hottest cluster cores.  Scd galaxies
\UGC8082/\NGC4848 and \AGC221406/\IC4040 have blue colors and detectable
remnants of \HI gas, while the other two undetected galaxies have quite red
\BI colors.  Spatially resolved \HI maps for \UGC8082/\NGC4848 (Bravo-Alfaro
\etal 2000) place the remaining \HI gas aligned directly along the major axis
and trailing the nontruncated side of the galaxy, which extends to 15 \hkpc
and 7 R$_d$, with a perturbed \HI gas distribution offset from the optical
centroid by a considerable 8 \hkpc.  The direction of the trailing wake of \HI
gas, and the correlation between \Halpha truncation and \HI stripping, may
support a radial orbit and edge-on infall path across the sky for this galaxy,
located on the north-west edge of the cluster diffuse X-ray flux (Vikhlinin,
Forman, \& Jones 1997).  The velocity offset of the galaxy from the cluster
redshift is low (273 \kms) for its radius of 582 \hkpc, which suggests a large
transverse velocity component of order 1,000 \kms and a corresponding
timescale of 10$^7$ h$^{-1}$ years since the onset of \HI gas stripping.  The
timescale for the disturbance of molecular gas, and subsequent star formation
suppression, on the truncated side of the disk would then be $\le$ 5 $\times$
10$^7$ h$^{-1}$ years, half of a full disk rotation.
However, Vollmer \etal (2001a) argue persuasively that this galaxy could not
have experienced the observed amount of \HI gas loss through ram pressure
stripping without having passed through the center of the cluster, and have
successfully simulated the distribution of the \HI gas and \Halpha flux
through re-accretion of atomic gas onto the disk.  (In their model, a
significant fraction of the stripped gas does not escape from the potential
when stripped but falls back onto the disk, triggering a burst of star
formation and explaining both the strong \Halpha flux and the blue colors.)

In contrast, \AGC221406/\IC4040 lies well with the projected diffuse X-ray gas
component of A1656 (Vikhlinin, Forman, \& Jones 1997).  Though the galaxy is
{\it asymmetric}, the maximum extent of \Halpha flux is 5.5 \hkpc, placing it
just above the predicted stripping radius for a galaxy of this size (see
Figure~\ref{fig:RstrRd}) and suggesting that gas stripping may have commenced
throughout all four disk quadrants.  The remaining \HI gas lies again along
the major axis (Bravo-Alfaro \etal 2000), supporting edge-on infall, though
the galaxy projected location deep within the X-ray gas at a radius of 250
\hkpc and the concentration of \HI gas along the more truncated \Halpha side
of the disk suggest that this galaxy first encountered the intracluster medium
more than a half rotational period ago (5 $\times$ 10$^7$ h$^{-1}$ years).
The large, positive velocity offset from the cluster redshift (+829 \kms), the
small offset on the sky (4 \hkpc) between the remaining \HI gas and the
optical centroid, and the orientation of the \Halpha rotation curve (the side
of maximum \Halpha extent rotates away from us, while the more truncated side
comes towards us from out of the sky) suggest that this galaxy is falling into
the cluster core from the foreground.  Given this orientation and history, we
cannot significantly constrain the timescales for gas stripping ($\ge$ 5
$\times$ 10$^7$ h$^{-1}$ years) or star formation suppression.

\AGC130204/\CGCG540-115, at a similarly small radius within A426, has fairly
weak \Halpha emission on the nontruncated side, which decreases by a factor of
five in strength and becomes quite patchy, where detectable, on the truncated
side (note, however, that \fnii6584\AA\ can be traced smoothly along both
sides of the entire optical disk).  \UGC8096/\IC3949 shows \Halpha in a
fascinating combination of emission on one side, and a truncated absorption
trough on the other, possibly a dynamic example of a later phase of the
transition from spiral to S0.  It is ranked as a post-starburst from its blue
spectral features (Caldwell \etal 1999), and VLA \HI observations
(Bravo-Alfaro \etal 2001) limit the \HI gas mass to $\le$ 3 $\times$ 10$^7$
M$_{\odot}$.

We examined the clusters which showed large number of {\it asymmetric} members
in detail, to see whether there were other candidate spiral galaxies in the
core regions which ought to show asymmetry from interaction with the
intracluster medium.  The remaining galaxies in the inner 750 \hkpc were
strongly \HI deficient and appeared to have truncated \Halpha on both sides of
the disk (see discussion in Section~\ref{subsubsec:Stripped}), indicating that
they had already passed through the intracluster medium.  Extending our search
out to 900 \hkpc, we note the case of \UGC2617, the sole hot cluster core
member offset significantly above the model stripping radius in
Figure~\ref{fig:RstrRd} which does not show signs of asymmetry.  A large Sc
galaxy located 858 \hkpc from the center of A426, \UGC2617 shows strong,
extended \Halpha emission on both sides of the disk despite its high \HI
deficiency (\HI$_{def}$ = 0.84).  At this large a clustercentric radius, it is
unlikely that the intracluster medium has caused the \HI gas loss, as a
first-pass galaxy would not yet have reached significant concentrations of hot
gas and a full pass through the intracluster medium would have truncated the
\Halpha emission for a galaxy of this morphological type; we suggest that its
\HI deficiency is unrelated to ram pressure stripping.

We further explored the remaining five A1656 galaxies within our dynamical
sample also observed by Bravo-Alfaro \etal (\AGC221206/\CGCG160-058, \UGC8118,
\AGC221409, \UGC8128/\NGC4911 and \UGC8140, all with {\it normal} \Halpha
distributions) for evidence of ram pressure stripping.  Four lie well to the
north of the cluster X-ray gas and show strong \HI gas, well-centered and
extending beyond the optical disks.  The fifth, \UGC8128/\NGC4911, located at
a radius of 375 \hkpc, is a massive spiral with M$_I - 5$ log h = -22.5;
Biviano \etal (1996) suggest that it is the dominant galaxy of a group which
recently passed through the cluster core.  The small offset between the
optical and (deficient) HI gas mass centroids of 2-3 \hkpc {\it perpendicular}
to the major axis, suggests a face-on infall path with gas stripping taking
effect across the entire disk simultaneously (see also the velocity map
derived in Amram \etal 1992).  In summary, none are viable candidates for
interaction with the intracluster medium and edge-on infall, suggesting that
our \Halpha asymmetry criterion is identifying such galaxies as exist within
the dynamical sample.

The {\it asymmetric} galaxies fall into two groups on the basis of optical
colors, with \BI colors either $<$ 1.5 (blue) or $>$ 2.0 (red).  The blue
galaxies have the small $B/T$ fractions of late type spirals, but are even
bluer in \BI, have intrinsically smaller lengths of R$_d$, R$_b$, and \Halpha
extent, and are profoundly \HI deficient -- suggestive of a central starburst
phase stimulated by the gas stripping process.  The red galaxies are larger
and brighter and have $B/T$ fractions higher than the mean of the early type
(Sa through Sbc) galaxies in the sample.  They have redder \BI colors, range
between normal and deficient amounts of \HI gas, and the lengths of R$_b$ and
the \Halpha extent are slightly less {\it in units of R$_d$}, evidence of mild
truncation of star formation in the outer regions of the disk.  Taken
together, the evidence supports a less extreme interaction with the
intracluster medium, buffered by a low impact parameter and/or by the strength
and shape of the potential well.  These groups appear to be in two parallel,
rather than sequential, phases along a single evolutionary path.

In summary, galaxies with asymmetric disk \Halpha flux, and a corresponding
decrease in \HI gas, can be found in a range of cluster environments, though
concentrated in the hot X-ray gas cores of rich clusters.  As these galaxies
will be edge-on rather than face-on to the intracluster medium through which
they fall, they should present a lower limit to the strength of the ram
pressure stripping on the disk.  \Halpha flux can be truncated on the side of
the disk where the \HI gas mass is slightly low but is not yet significantly
depleted, indicating that the star formation is quite sensitive and responsive
to changes in the localized gas mass.  Both the red and the blue populations
span the full range of \HI deficiency, suggesting that the global
instantaneous star formation rate is not a strong function of the available
\HI gas mass (in agreement with Bothun, Schommer, \& Sullivan 1982, for
cluster spirals, and as found for our total sample of normal spirals), and in
fact the \BI color correlates most strongly with spiral type.

The gas stripping could be caused by tidal forces, either from interaction
with the cluster potential as the galaxy passes through the central core or
from close (within 10 \hkpc) galaxy-galaxy interactions with other infalling
spirals (Gnedin 2003a; also Abadi, Moore, \& Bower 1999 whose models indicate
that ram pressure stripping alone is insufficient to generate the observed
stripping in larger spirals).  However, the strong correlation between the
orientation of the truncated side of the disk and the vector to the cluster
core, the strong stripping of gas from the outer portions of the optical disk,
and the spatial bias towards inner 600 \hkpc of the cluster (the maximum
radius at which the hot gas should begin to strip gas within the richest
clusters), and the lack of distortion in \Ib disk surface brightness profiles
suggests that ram pressure stripping is an important mechanism for suppressing
star formation in these disks.  The timescale for a full disk rotation, after
which all quadrants of the disk would have rotated through the leading,
exposed edge, is roughly one-tenth of a single pass through the intracluster
medium at the core of a rich cluster.  Coupled with the decrease in \Halpha
flux and \HI gas stripped from one side of the disk but no significant
distortion of the disk in the \Ib, this suggests that the stripping occurs on
timescales well under 10$^8$ years.  This is in agreement with predictions of
star formation lifetimes based on star formation rates and \HI gas masses
(Kennicutt \etal 1984); note also that Abadi, Moore, \& Bower (1999) predict
that ram pressure stripping of \HI gas will operate relatively quickly on a
timescale of 10$^7$ years, significantly faster than galaxy-galaxy tidal
interactions.  The incidence of {\it asymmetric} galaxies, and timescale
estimates derived above, suggest a current mass accretion rate of 100 h$^3$
M$_{\odot}$/year for the richest clusters. This is strikingly similar to the
gas condensation rate from the intracluster medium, estimated from the
measured excess central emission in cooling flow clusters (\eg Lufkin,
Sarazin, \& White 2000).

\subsection{Stripped Spirals}
\label{subsubsec:Stripped}

What happens to {\it asymmetric} spirals after gas has been stripped from all
disk quadrants?  Their next evolutionary phase should last longer, as the
abrupt suppression of wide-scale young star formation is followed by a slower
decrease in the remaining star formation concentrated in the protected inner
region ($\le$ 5 \hkpc) of the disk.  Gas may not have been removed from this
inner region, but the draining of the external reservoir insures that there
will be little additional fuel funneled inwards in the future.  Galaxies
within this phase should be characterized by (a) young star formation confined
to the inner few \hkpc across the entire disk, (b) \HI deficiency, (c) warm
cluster membership, and (d) a more relaxed orbital distribution than the {\it
asymmetric} galaxies.

A fraction of infalling field spirals will exhibit these characteristics
without passing through an {\it asymmetric} phase.  Recent numerical
simulations (Gnedin 2003a, 2003b; Moore \etal 1998) suggest that tidal forces,
from a time-varying cluster potential or from galaxy-galaxy interactions
(harassment), may be as important as ram pressure stripping in transforming
spirals in cluster cores, and may operate more efficiently within infalling
groups in the outer regions of clusters.  Large spirals are predicted to
experience halo truncation (beyond the optical radius), vertical heating
resulting in a thickened disk, and tidal shocks leading to gas dissipation.
The gas can lose enough angular momentum that it sinks to the galaxy nucleus
(Barnes \& Hernquist 1996) and fuels a starburst phase.  In summary, face-on
infall paths will strip the gas across the entire outer region of the disk
simultaneously, more efficiently than edge-on interactions (Abadi, Moore, \&
Bower 1999), and mechanisms such as harassment or tidal interactions can
remove the gas reservoir, on longer timescales.

We identify a candidate {\it stripped} population from the full dynamic
sample, searching for galaxies with truncated star formation across the entire
disk.  These galaxies have a maximum extent of \Halpha flux of less than 5
\hkpc and of less than 3 $R_d$.  Most are strongly deficient in \HI gas.  They
are found in a more relaxed orbital distribution than the {\it asymmetric}
galaxies, out to 2 \hMpc from the cores of a wider range of clusters.  In
combination with their early morphological types and slightly reddened \BI
colors, this suggests that they are comprised of galaxies undergoing
non-catastrophic stripping (\ie no starburst phase, and still identifiable as
spirals), and of formerly {\it asymmetric} galaxies which have penetrated the
intracluster medium and continue along their orbital paths in a post-starburst
phase.

Like the {\it asymmetric} galaxies, the {\it stripped} spirals were identified
within the dynamical sample on the basis of \Halpha extent.  The primary
criterion is the truncation of \Halpha extent to within 5 \hkpc; the second
limitation to within 3 $R_d$ was added to eliminate the inclusion of
unstripped galaxies with normal amounts of \HI gas but with intrinsically
small disk sizes.  The {\it stripped} spirals are distributed along the
extreme lower edge of the spatial distribution for all spirals, where \HI
deficient galaxies within the cores of hot clusters are located; relaxing the
criteria would quickly result in the inclusion of galaxies which have normal
amounts of \HI gas, and are located in the field and in cooler clusters.

\HI normal galaxies across the full range of environments within the dynamical
sample lie within the same region well above the predicted stripping radius.
The exceptions are the hot cluster core members (encircled on
Figure~\ref{fig:RstrRd}), which follow the lower edge of the envelope.  For
early type, \HI deficient galaxies there is a clear distinction between those
located beyond the warm cluster cores well away from the intracluster medium,
which still lie above the stripping radius, and those within the warm and hot
cores, for which \Halpha is truncated to below the stripping radius.  This
environmental distinction is not reproduced for late type, HI deficient
spirals.

Though the specific {\it stripped} criteria were motivated by the distribution
of \Halpha flux throughout the dynamical sample, they agree fairly closely
with model predictions which balance ram pressure stripping forces against the
gravitational restoring force of the galaxy potential.
Figure~\ref{fig:RstrRd} shows good agreement between the two approaches, and
the distribution of galaxies through the dynamical sample as a function of
morphological type and \HI gas deficiency.  The extreme paucity of late type
spirals within warm clusters is highlighted, and the lack of late type {\it
stripped} galaxies suggests that infalling late type spirals are
morphologically altered beyond recognition (Moore \etal 1996) or that star
formation suppression in the outer regions of the disk leads to an early type
classification (Koopmann 1997).  There is a significant population of early
type spirals within clusters of all temperatures.  Those found within warm
clusters follow the field and cold cluster distribution when their gas
reservoirs are intact, though those within the hottest cluster cores fall
along the lower edge.  There is a clear distinction for \HI deficient galaxies
however, as the two populations begin to separate.  The locus of cool cluster
galaxies does not shift significantly, but those within warm clusters exhibit
less extended \Halpha flux and tend to fall below the stripping radius.  Early
type galaxies within warm clusters are the only population found in this
regime.

The {\it stripped} galaxies have a moderately tight radial distribution, more
relaxed than the {\it asymmetric} galaxies but still fairly localized to the
warm cluster cores, for which a crossing time is 10$^9$ years.  This is a
similar timescale to that of the {\it stripped} phase, given predictions from
stellar population models and the strength of the Balmer features in spectra
of post-starburst galaxies (1.5 $\times$ 10$^9$ years, \cf Poggianti \&
Barbaro 1996).  One would thus expect 10 to 50 times more early type galaxies
to appear in the {\it stripped} phase than the {\it asymmetric}, based on
timescales alone and considering that some infalling galaxies will not pass
through the {\it asymmetric} stage (\eg face-on infall paths).  The fading
expected in broadband colors is not sufficient to preferentially remove {\it
stripped} galaxies from within our observational sample, but a significant
decrease in blue light diameter R$_b$ could force some below 30\arcsec\ in
size (see discussion in \pthree), accounting for the relative counts of less
than one to three.

\subsection{Quenched Spirals}

What happens to {\it stripped} galaxies after the remnants of young star
formation along the entire disk slow to a halt, and there are no reservoirs of
gas available for replenishment?  Galaxy morphology, and star formation
properties, should become dominated by the older, underlying stellar
population.  Our key tracer of star formation, \Halpha emission along the
disk, fades into obscurity, to be replaced by an emerging \Halpha absorption
feature characteristic of a stellar population dominated by A-type stars.  We
return to our complete dynamical sample, and examine the \Halpha flux
characteristics throughout for evidence of such features.  While the majority
(93\%) of the sampled galaxies exhibit strong \Halpha emission, the remaining
21 galaxies are characterized by a strong, extended absorption feature.  We
define them as {\it quenched} spirals.  The \Halpha absorption trough is deep
and can easily be traced through the nucleus and along the disk.  It extends
to a radius at or beyond 2 $R_d$ for all but three of the galaxies.  The {\it
quenched} galaxies are extremely \HI deficient (see Table~\ref{tab:NASQ}), and
$\frac{2}{3}$ have no detected \HI gas at all.

{\it Quenched} spiral bulge fractions range from 20\% to 60\% and inclination
angles from 45$^{\circ}$\ to 90$^{\circ}$, falling within the envelope of the
complete dynamical sample.  Nine of these galaxies are listed in the \UGC
(Nilson 1973), and for the remainder type codes were taken from another
catalog (\NGC, Dreyer 1888; \IC, Dreyer 1895; \CGCG, Zwicky \etal 1968) if
available and determined from a visual examination of the POSS plates for the
two previously uncataloged galaxies, before inclusion into the observing
sample.  Assigned type codes range from Sa through Sc, but are predominantly
early types.  Note the difficulty of typing the most edge-on portion of the
sample; it can become problematical to assess the level of spiral arm
structure, though one can look for dust lanes and signs of extinction.

We find the {\it quenched} spirals preferentially in the rich clusters A1656,
A426 and A1367; the remaining half lie in the poorer clusters and none are
found in Cancer or N507.  Only one (\UGC11633, typed Sa) is from the field
sample.  These galaxies are situated as far out as a radius of 2.5 \hMpc,
quite beyond the cluster cores where spirals encounter both the intracluster
medium and the enhanced tidal effects from close encounters and the cluster
potential.  This does not necessarily contradict formation through these
mechanisms, however, as the very galaxy-galaxy interactions which lead to gas
loss may also be responsible for redirecting the galaxies onto highly
eccentric and loosely bound orbits (Balogh, Navarro, \& Morris 2000).  Their
extreme \HI deficiency indicates that the entire galaxy has been stripped,
rather than just the outer regions of the disk.  This suggests that tidal
interactions, rather than ram pressure stripping which operates most
efficiently in the outer disk, may be the key mechanism.

The simplest interpretation of the {\it quenched} galaxies is that they lie at
a later stage along the transition from infalling field spirals to cluster
S0s, passing or having already passed through a cluster core at least
once. The dynamical sample was selected for spiral appearance with no
deliberate inclusion of S0s.  Any S0 contamination would thus come in the form
of those which most closely resemble spirals rather than more spheroidal
systems, and be the most likely candidates for reformed spirals within the S0
population.  It is also clear that the {\it quenched} population cannot all be
misclassified S0s, due to the cases of clear spiral arm morphology (\eg
\UGC1350 within Figure~\ref{fig:RCHISample}).  The relaxed orbital
distribution, as shown in Figure~\ref{fig:QuenS0Rad}, is completely different
from that of the tightly centered S0 population.

\begin{figure} [htbp]
  \begin{center}\epsfig{file=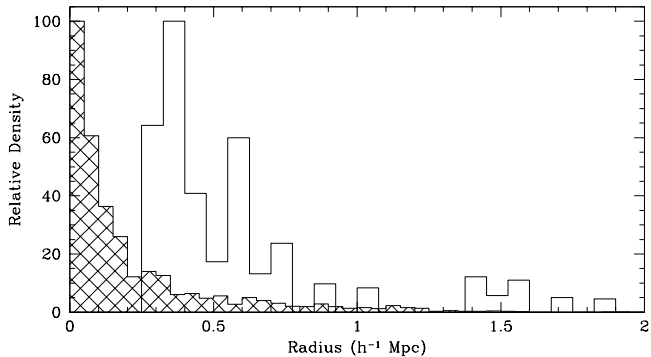,width=3.4truein}\end{center}
  \caption[S0 and quenched radial histogram]
{Radial distribution of {\it quenched} spirals (open) and of S0s
(crossed), as a function of clustercentric radius.  We find no {\it quenched}
spirals within the inner 200 \hkpc of any of the clusters, where the
S0 distribution peaks, and a Kolmogorov-Smirnov test estimates the
probability of the two groups being drawn from different parent populations at
greater than 96\%.}
  \label{fig:QuenS0Rad} 
\end{figure}

These galaxies share certain characteristics with the population of E+A
galaxies found in intermediate redshift clusters (Dressler \& Gunn 1983) and
locally in the field (Zabludoff \etal 1996) or post-starburst (PSB) galaxies
in local clusters (Caldwell, Rose, \& Dendy 1999).  These designate galaxies
which are characterized by (1) strong Balmer absorption lines, indicating an
older stellar population of A-type stars and recent star formation, coupled
with (2) an absence of emission lines (specifically, \foii3727\AA) which rules
out current star formation.  Stellar population models suggest a recently
ended ($\sim$ 1 Gigayear) burst of star formation, rather than the more
constant rate of a typical spiral galaxy (Leonardi \& Rose 1996).  The finding
of a significant population of field E+As, with evidence of tidal effects
(Zabludoff \etal 1996), implies that these galaxies can manifest purely
through galaxy-galaxy interactions.  Their presence in rich clusters could
then be attributed to previous interactions within an infalling group which
left a signature, to tidal interactions within the cluster, or to
cluster-specific mechanisms (\eg ram pressure stripping) if there exists more
than one causative mechanism.

We lack the blue spectral coverage of \foii3727\AA ~and \Hbeta, \Hgamma and
\Hdelta necessary to make a full comparison with E+A spectra, our spectra
being confined to a fairly narrow band around \Halpha, and our broadband
coverage being restricted to \BI colors.  More importantly, we have selected
for spiral galaxies while these samples have been focused upon elliptical and
S0s (though Caldwell \etal 1996 find E+A galaxies which are determined to be
disk systems).  We have also selected highly inclined galaxies, in order to
determine velocity widths, while the Caldwell \etal (1996) sample tends
strongly towards face-on galaxies for which the structure and stellar
populations can be more easily examined.  Though we both have surveyed the
Coma cluster, there is thus little overlap (though see discussion of
\UGC8096/\IC3949 in Section~\ref{subsubsec:Asymmetrics}) between our
observational samples.  Nonetheless, it is possible that our quenched spirals
represent an evolved form of late type E+A galaxies, as the bulk of the PSB
galaxies do for early types.

\section{Conclusions}

A strong inverse relation is obtained between the fraction of spirals and that
of S0s, within the inner 1 \hMpc or including infalling groups of spirals
within up to 2 \hMpc in local cluster membership calculations.  The elliptical
fraction, in contrast, holds fairly constant across the full range of three
orders of magnitude in cluster X-ray luminosity.
We have explored the strong correlation between \HI gas stripping and the
consequential suppression of young star formation, finding a correlation
between the distribution of \HI flux and of \HII regions within the galaxy
disks.  To this end, we have divided the sample into four groups, on the basis
of \Halpha emission properties.

{\bf Group I}: {\it normal} spirals, with no particularly striking properties
in the extent and strength of the observed \Halpha emission flux.  Many, but
not all, of these spirals have the expected amount of atomic gas for their
optical size and morphological type.  Most of the field spirals fall into this
category, and the cluster spirals which do so tend to have similar properties
as the field galaxies.  This group also includes a small number of spirals
located in the cores of rich clusters with patchy, somewhat decremented
\Halpha emission flux extent.  They appear to be infalling at a face-on
orientation into the intracluster medium, and the gas reservoir is being
stripped simultaneously across the entire radial extent of the disk.

{\bf Group II}: {\it asymmetric} spirals, where the radial extents of \Halpha
emission traced on each side of the disk either differ by more than 5 \hkpc or
form a ratio of less than 1:2.  Galaxies with this degree of asymmetry are not
observed beyond 1 \hMpc in any of the clusters, and are found predominantly in
the richest cluster cores.  Half are deficient in \HI, with an upper limit of
30\% of the expected \HI gas, and the distribution of detected \HI gas
correlates with that of the \Halpha emission.  They are oriented
preferentially edge-on to the cores, with truncation of \Halpha flux {\it and}
\HI flux along the leading edge, suggesting that ram pressure stripping from a
first pass through the intracluster medium plays an important role in
generating this effect.  The suppression of star formation along the disk
occurs on a timescale similar to that of the \HI gas stripping, as we find
many galaxies where the gas stripping process has begun (low \HI gas content
and asymmetry in the two-horn \HI line profile) but not yet been completed (a
substantial amount of \HI gas remains) where star formation has already been
terminated along the leading edge of the disk.

The {\it asymmetric} galaxies fall into two groups, with \BI colors either $<$
1.5 (blue) or $>$ 2.0 (red).  The blue galaxies have the small $B/T$ fractions
of late type spirals, but are even bluer in \BI, have intrinsically smaller
lengths of R$_d$, R$_b$, and \Halpha extent, and are profoundly \HI deficient
-- suggestive of a central starburst phase stimulated by the gas stripping
process.  The red galaxies are larger and brighter and have $B/T$ fractions
higher than the mean of the early type (Sa through Sbc) galaxies in the
sample.  They have redder \BI colors, range between normal and deficient in
\HI, and the lengths of R$_b$ and the \Halpha extent are slightly less {\it in
units of R$_d$}, evidence of mild truncation of star formation in the outer
regions of the disk.  Taken together, the evidence supports a less extreme
interaction with the cluster, supported perhaps by a low impact parameter or
by the greater resistive force of a more massive potential well.  These two
groups appear to be on parallel, rather than sequential, tracks in their
morphological histories.

{\bf Group III}: {\it stripped} spirals, where the extent of \Halpha emission
is less than 5 \hkpc on both sides of the disk and extends to less than 3 disk
scale lengths.  The bulk of these galaxies are strongly deficient in \HI.
They are found in a more relaxed orbital distribution than the {\it
asymmetric} galaxies, out to 2 \hMpc from the cores of a wider range of
clusters.  In combination with their slightly reddened colors, this suggests
that they are comprised of less massive systems currently undergoing
non-catastrophic stripping (\ie no starburst phase, and still identifiable as
spirals), and of blue {\it asymmetric} galaxies which have already passed
through the cores and are now in a post-starburst phase.

{\bf Group IV}: {\it quenched} spirals, for which star formation has been
halted across the entire disk and \Halpha is found only in absorption.  These
galaxies range in appearance between possible edge-on S0s, and (primarily
early type) spirals with clear spiral arm structure.  They are 1 magnitude
fainter in \Ib than early type \HI normal field spirals, 0.5 magnitudes redder
in \BI, their disk scale lengths are a factor of 2 smaller, and the extent of
the stellar \Halpha (absorption) flux is quite truncated along the disk
(relative to the normal \Halpha extent).  \HI observations place an upper
limit of one-tenth of the expected \HI flux on the sample, and 90\% are
undetected.  These galaxies may serve to illustrate the transition stage of a
morphological transformation between infalling field spiral and cluster S0s.
Their current orbital distribution is far less radially concentrated than that
of present day S0s, and on a timescale of a few Gigayears they may slowly
blend with that population.

In the richest, hottest clusters we find primarily {\it asymmetric, stripped,}
and {\it quenched} galaxies, and {\it normal} edge-on infalling spirals within
1 \hMpc of the cores.  Less rich clusters contain primarily {\it normal}
spirals, at all radii.  This is consistent with a picture of infalling spirals
being significantly altered by ram-pressure stripping in the hot cores, while
galaxy-galaxy interactions and tidal forces play a role throughout the entire
cluster distribution, and at larger radii.

In summary, we have explored the formation and evolution of spiral galaxies in
local clusters through a combination of optical and \HI properties.  We find a
clear relationship between \HI gas stripping and the consequential suppression
of young star formation; both occur quickly within spirals infalling into an
intracluster medium of hot gas.  We have traced galaxies through a progression
of infall stages, beginning with infalling field spirals and transforming via
gas stripping and passive fading into cluster proto-S0s (not with a bang but a
whimper).

\section{Acknowledgments}

The data presented in this paper are based upon observations carried out at
the Arecibo Observatory, which is part of the National Astronomy and
Ionosphere Center (NAIC), at Green Bank, which is part of the National Radio
Astronomy Observatory (NRAO), at the Kitt Peak National Observatory (KPNO),
the Palomar Observatory (PO), and the Michigan--Dartmouth--MIT Observatory
(MDM). NAIC is operated by Cornell University, NRAO by Associated
Universities, inc., KPNO and CTIO by Associated Universities for Research in
Astronomy, all under cooperative agreements with the National Science
Foundation. The MDM Observatory is jointly operated by the University of
Michigan, Dartmouth College and the Massachusetts Institute of Technology on
Kitt Peak mountain, Arizona. The Hale telescope at the PO is operated by the
California Institute of Technology under a cooperative agreement with Cornell
University and the Jet Propulsion Laboratory.  We thank the staff members at
these observatories who so tirelessly dedicated their time to insure the
success of these observations.

We also thank Sc project team members John Salzer, Gary Wegner, Wolfram
Freudling, Luiz da Costa, and Pierre Chamaraux, and also Shoko Sakai and Marco
Scodeggio, for sharing their data in advance of publication.  Mario Abadi
kindly provided the output data from a ram-pressure stripping model, and Danny
Dale provided tabulated data in electronic form.  N.P.V. is pleased to thank
Ann Zabludoff and Sandra Faber for many enlightening discussions regarding
cluster formation, and Karl Gebhardt for generously sharing his adaptive
kernel density fitting routines.  
The text of this manuscript was much improved by careful reading on the part
of Richard Ellis.  We thank the anonymous referee for helpful comments on the 
manuscript.

N.P.V. is a Guest User, Canadian Astronomy Data Center, which is operated by
the Dominion Astrophysical Observatory for the National Research Council of
Canada's Herzberg Institute of Astrophysics.  This research has made use of
the NASA/IPAC Extragalactic Database (NED) which is operated by the Jet
Propulsion Laboratory, California Institute of Technology, under contract with
NASA, and NASA's Astrophysics Data System Abstract Service (ADS).  This
research was supported by NSF grants AST92--18038 and AST95--28860 to
M.P.H. and T.H., AST90--23450 to M.P.H., AST94--20505 to R.G., and
NSF--0123690 via the ADVANCE Institutional Transformation Program at NMSU, and
NASA grants GO-07883.01-96A to N.P.V. and NAS5--1661 to the WFPC1 IDT.
N.P.V. acknowledges the generous support of an Institute of Astronomy rolling
grant from PPARC, reference number PPA/G/O/1997/00793.


\end{document}